\documentclass[preprint2, twocolappendix, times, tighten, appendixfloats]{aastex6}

\newcommand{\hst}{{\it HST}}
\newcommand{\spitzer}{{\it Spitzer}}
\newcommand{\chandra}{{\it Chandra}}

\shorttitle{Cluster Galaxy Star Formation Evolution}
\shortauthors{Wagner et al.}

\begin{document}
\title{The Evolution of Star Formation Activity in Cluster Galaxies Over $0.15<\lowercase{z}<1.5$}

\author{Cory R. Wagner\altaffilmark{1,6}, St\'ephane Courteau\altaffilmark{1}, Mark Brodwin\altaffilmark{2}, S. A. Stanford\altaffilmark{3}, Gregory F. Snyder\altaffilmark{4}, and Daniel Stern\altaffilmark{5}}
\altaffiltext{1}{Department of Physics, Engineering Physics \& Astronomy, Queen's University, 64 Bader Lane, Kingston, Ontario, Canada K7L 3N6}
\altaffiltext{2}{Department of Physics and Astronomy, University of Missouri, 5110 Rockhill Road, Kansas City, MO 64110, USA}
\altaffiltext{3}{Department of Physics, University of California, One Shields Avenue, Davis, CA 95616, USA}
\altaffiltext{4}{Space Telescope Science Institute, 3700 San Martin Drive, Baltimore, MD 21218, USA}
\altaffiltext{5}{Jet Propulsion Laboratory, California Institute of Technology, Pasadena, CA 91109, USA}
\altaffiltext{6}{\href{mailto:cory.wagner@queensu.ca}{cory.wagner@queensu.ca}}

\begin{abstract}
We explore 7.5 billion years of evolution in the star formation activity
of massive ($M_{\star}>10^{10.1}\,M_{\odot}$) cluster galaxies using
a sample of 25 clusters over $0.15<z<1$ from the Cluster Lensing
And Supernova survey with \textit{Hubble} and 11 clusters over $1<z<1.5$
from the IRAC Shallow Cluster Survey. Galaxy morphologies are determined
visually using high-resolution \textit{Hubble Space Telescope} images.
Using the spectral energy distribution fitting code CIGALE, we measure
star formation rates, stellar masses, and 4000 \AA\ break strengths.
The latter are used to separate quiescent and star-forming galaxies
(SFGs). From $z\sim1.3$ to $z\sim0.2$, the specific star formation
rate (sSFR) of cluster SFGs and quiescent galaxies decreases by factors
of three and four, respectively. Over the same redshift range, the
sSFR of the entire cluster population declines by a factor of 11,
from $0.48\pm0.06\;\mathrm{Gyr}^{-1}$ to $0.043\pm0.009\;\mathrm{Gyr}^{-1}$.
This strong overall sSFR evolution is driven by the growth of the
quiescent population over time; the fraction of quiescent cluster
galaxies increases from $28^{+8}_{-19}\%$ to $88^{+5}_{-4}\%$ over $z\sim1.3\rightarrow0.2$. The majority of the growth occurs at $z\gtrsim0.9$, where the quiescent fraction increases by 0.41. While
the sSFR of the majority of star-forming cluster galaxies is at the
level of the field, a small subset of cluster SFGs have low field-relative
star formation activity, suggestive of long-timescale quenching. The
large increase in the fraction of quiescent galaxies above $z\sim0.9$,
coupled with the field-level sSFRs of cluster SFGs, suggests that
higher redshift cluster galaxies are likely being quenched quickly.
Assessing those timescales will require more accurate stellar population
ages and star formation histories.
\end{abstract}


\section{Introduction}
\label{sec:introduction}

It is well known that the star formation (SF) in cluster galaxies
in the $z\lesssim1$ regime decreases with the age of the Universe
\citep{couch1987,saintonge2008,finn2008,webb2013}, while the cluster
early-type galaxy (ETG) fraction increases \citep{stanford1998,poggianti2009}.
By $z\sim1$, a correlation between SF and environment appears to
be in place in galaxy clusters \citep{muzzin2012}, with high fractions
of quiescent galaxies relative to the field leading to significant
differences between the stellar mass functions of field and cluster
galaxies \citep{vanderburg2013}. The $1<z<1.5$ epoch does reveal
an era of substantial SF activity for some cluster galaxies \citep{brodwin2013,zeimann2013,alberts2014}.
However, the bulk of studies investigating cluster SF at all redshifts
address mostly the total SF activity within a given cluster, with
no attempt to quantify morphological dependencies. While this allows
for larger sample sizes, thus reducing random uncertainties, it smooths
over any potential differences that may exist amongst subpopulations
within the cluster.

\citet[hereafter Paper I]{wagner2015} focused on quantifying the
SF activity of massive cluster galaxies of differing morphologies
over $1<z<1.5$ and found that ETGs, with mean star formation rates
(SFRs) of $\sim$$7\,M_{\odot}\,\mathrm{yr}^{-1}$, contribute 12\%
of the vigorous SF in clusters at that redshift range. This high and
consistent SF activity, coupled with an increasing fraction of ETGs
from $z=1.5$ to 1.25 was taken as evidence of major mergers driving
mass assembly in clusters at $z\sim1.4$. At later times, cluster
galaxies appear to be transitioning away from an epoch of enhanced
SF activity to one of steady quenching (the depletion of the cold
gas necessary for the formation of new stars).

A number of possible quenching mechanisms, such as strangulation \citep{larson1980},
ram-pressure stripping \citep{gunn1972}, and active galactic nucleus
(AGN) feedback, are at play in clusters, acting over different timescales.
Strangulation is a long-timescale (several Gyr) process that removes
hot, loosely bound gas from a galaxy's halo, gas that could otherwise
cool and fall onto the galaxy. The in situ cold gas, however, is not
removed through this process, and the galaxy can continue to form
new stars until this supply is depleted. Ram-pressure stripping, where
the dense intracluster medium strips a galaxy's cold interstellar
gas as it travels through the cluster, occurs typically on a $\sim$1
Gyr timescale. AGN feedback, the heating and/or expelling of cold
gas from the core of a galaxy, is a process that can occur on timescales
as short as a few hundred million years \citep{dimatteo2005,hopkins2006},
and can be a byproduct of major merger activity \citep{springel2005}.

In this work, we extend our study of cluster SF activity to the full
redshift range $0.15<z<1.5$, using the Cluster Lensing and Supernova
survey with \textit{Hubble} \citep[CLASH;][]{postman2012} and IRAC
Shallow Cluster Survey \citep[ISCS;][]{eisenhardt2008} cluster samples.
CLASH covers the range $0.15<z<1$, while our subset of ISCS spans
$1<z<1.5$. Measuring the SF activity of cluster galaxies over a large
range in cosmic times may provide insight into the dominant quenching
mechanism(s) at different redshifts.

Our paper is laid out as follows. A description of our cluster samples
is found in Section\ \ref{sec:cluster_sample}, while the galaxy
sample selection and estimates of stellar masses ($M_{\star}$) and
SFRs through spectral energy distribution (SED) fitting are presented
in Section\ \ref{sec:sample_selection}. Section\ \ref{sec:results}
explores and discusses cluster galaxy SF activity as functions of
stellar mass, redshift, and galaxy morphology. We summarize our results
in Section\ \ref{sec:conclusion}. We adopt a WMAP7 cosmology \citep{komatsu2011},
with ($\Omega_{\Lambda}$, $\Omega_{M}$, $h$) = (0.728, 0.272, 0.704).
Cluster halo mass measurements and uncertainties compiled from the
literature are scaled to this cosmology, assuming the respective authors
have followed the conventions described by \citet{croton2013} with
respect to the treatment of $h$.

\section{Cluster Samples}

\label{sec:cluster_sample}

In Table\ \ref{tab:clusterList}, we provide relevant details of
our cluster sample, including cluster names, positions, spectroscopic
redshifts, and halo mass and velocity dispersion measurements compiled
from the literature. While we list the full cluster names in tables,
for brevity we use shortened versions for CLASH clusters in the subsequent
text.

\begin{deluxetable*}{lccccccccccccc}
\tablecaption{CLASH and ISCS Cluster Samples\label{tab:clusterList}}
\tabletypesize{\footnotesize}
\tablecolumns{14}
\tablehead{
\colhead{Cluster} &
\colhead{R.A.} &
\colhead{Declination} &
\colhead{$z$} &
\colhead{$M_{200}$} &
\colhead{$\sigma_{\mathrm{vel}}$} &
\multicolumn{2}{c}{Reference} &
\colhead{} &
\multicolumn{5}{c}{Final Sample Galaxy Counts} \\
\cline{7-8}
\cline{10-14}
\colhead{} &
\colhead{(J2000)} &
\colhead{(J2000)} &
\colhead{} &
\colhead{($10^{14}\,M_{\odot}$)} &
\colhead{($\mathrm{km}\,\mathrm{s}^{-1}$)} &
\colhead{$M_{200}$} &
\colhead{$\sigma_{\mathrm{vel}}$} &
\colhead{} &
\colhead{$N_{\mathrm{tot}}$} &
\colhead{$N_{\mathrm{ETG}}$} &
\colhead{$N_{\mathrm{LTG}}$} &
\colhead{$N_{\mathrm{SFG}}$} &
\colhead{$N_{\mathrm{Q}}$}
}
\startdata
\sidehead{CLASH}
Abell 383 & 02:48:03.36 & $-$03:31:44.7 & 0.187 & $11.3\pm3.8$ & $931\pm59$ & 1 & 6 &  & 28 & 25 & 3 & 2 & 26\\
Abell 209 & 01:31:52.57 & $-$13:36:38.8 & 0.209 & $21.9\pm4.9$ & $1320_{-67}^{+64}$ & 1 & 7 &  & 21 & 18 & 3 & 3 & 18\\
Abell 1423 & 11:57:17.26 & $+$33:36:37.4 & 0.213 & $7.2\pm1.7$ & $759_{-51}^{+64}$ & 2 & 8 &  & 16 & 13 & 3 & 5 & 11\\
Abell 2261 & 17:22:27.25 & $+$32:07:58.6 & 0.224 & $32.8\pm7.4$ & $780_{-60}^{+78}$ & 1 & 8 &  & 27 & 24 & 3 & 2 & 25\\
RXJ 2129.7$+$0005 & 21:29:39.94 & $+$00:05:18.8 & 0.234 & $8.7\pm2.5$ & $858_{-57}^{+71}$ & 1 & 8 &  & 28 & 26 & 2 & 2 & 26\\
Abell 611 & 08:00:56.83 & $+$36:03:24.1 & 0.288 & $22.4\pm6.4$ & $\cdots$ & 1 & $\cdots$ &  & 36 & 33 & 3 & 5 & 31\\
MS 2137$-$2353 & 21:40:15.18 & $-$23:39:40.7 & 0.315 & $19.3\pm7.5$ & $\cdots$ & 1 & $\cdots$ &  & 14 & 8 & 6 & 1 & 13\\
RXJ 2248.7$-$4431 & 22:48:44.29 & $-$44:31:48.4 & 0.348 & $26.7\pm9.5$ & $1660_{-135}^{+200}$ & 1 & 9 &  & 39 & 37 & 2 & 3 & 36\\
MACS 1931.8$-$2635 & 19:31:49.66 & $-$26:34:34.0 & 0.352 & $21.7\pm10.1$ & $\cdots$ & 1 & $\cdots$ &  & 14 & 11 & 3 & 0 & 14\\
MACS 1115.9$-$0129 & 11:15:52.05 & $+$01:29:56.6 & 0.353 & $23.7\pm5.5$ & $\cdots$ & 1 & $\cdots$ &  & 24 & 21 & 3 & 2 & 22\\
RXJ 1532.9$+$3021 & 15:32:53.78 & $+$30:20:58.7 & 0.363 & $8.5\pm3.3$ & $\cdots$ & 1 & $\cdots$ &  & 26 & 23 & 3 & 5 & 21\\
MACS 1720.3$+$3536 & 17:20:16.95 & $+$35:36:23.6 & 0.391 & $20.6\pm6.1$ & $\cdots$ & 1 & $\cdots$ &  & 41 & 34 & 7 & 8 & 33\\
MACS 0416.1$-$2403 & 04:16:09.39 & $-$24:04:03.9 & 0.396 & $15.3\pm3.7$ & $996_{-36}^{+12}$ & 1 & 10 &  & 78 & 62 & 16 & 16 & 62\\
MACS 0429.6$-$0253 & 04:29:36.00 & $-$02:53:09.6 & 0.399 & $13.9\pm5.0$ & $\cdots$ & 1 & $\cdots$ &  & 28 & 24 & 4 & 4 & 24\\
MACS 1206.2$-$0847 & 12:06:12.28  & $-$08:48:02.4 & 0.440 & $25.8\pm6.0$ & $1042_{-53}^{+50}$ & 1 & 11 &  & 74 & 60 & 14 & 16 & 58\\
MACS 0329.7$-$0211 & 03:29:41.68 & $-$02:11:47.7 & 0.450 & $12.3\pm2.8$ & $\cdots$ & 1 & $\cdots$ &  & 67 & 62 & 5 & 6 & 61\\
RXJ 1347.5$-$1145 & 13:47:30.59 & $-$11:45:10.1 & 0.451 & $48.7\pm12.5$ & $1163\pm97$ & 1 & 12 &  & 37 & 35 & 2 & 7 & 30\\
MACS 1311.0$+$0310 & 13:11:01.67 & $-$03:10:39.5 & 0.494 & $6.5\pm0.4$ & $\cdots$ & 3 & $\cdots$ &  & 37 & 32 & 5 & 8 & 29\\
MACS 1149.6$+$2223 & 11:49:35.86 & $+$22:23:55.0 & 0.544 & $35.5\pm7.9$ & $1840_{-170}^{+120}$ & 1 & 13 &  & 108 & 91 & 17 & 23 & 85\\
MACS 1423.8$+$2404 & 14:23:47.76 & $+$24:04:40.5 & 0.545 & $8.1\pm1.4$ & $1300_{-170}^{+120}$ & 3 & 13 &  & 54 & 41 & 13 & 11 & 43\\
MACS 0717.5$+$3745 & 07:17:31.65 & $+$37:45:18.5 & 0.548 & $38.0\pm7.6$ & $1612\pm70$ & 3 & 14 &  & 132 & 106 & 26 & 29 & 103\\
MACS 2129.4$-$0741 & 21:29:25.32 & $-$07:41:26.1 & 0.570 & $\cdots$\tablenotemark{a} & $1400_{+200}^{+170}$ & $\cdots$ & 13 &  & 57 & 49 & 8 & 15 & 42\\
MACS 0647.8$+$7015 & 06:47:50.03 & $+$70:14:49.7 & 0.584 & $19.7\pm6.0$ & $900_{-180}^{+120}$ & 1 & 13 &  & 38 & 36 & 2 & 5 & 33\\
MACS 0744.9$+$3927 & 07:44:52.80 & $+$39:27:24.4 & 0.686 & $25.6\pm7.0$ & $1110_{-150}^{+130}$ & 1 & 13 &  & 49 & 39 & 10 & 14 & 35\\
CLJ 1226.9$+$3332 & 12:26:58.37 & $+$33:32:47.4 & 0.890 & $22.2\pm1.4$ & $1143\pm162$ & 3 & 15 &  & 61 & 50 & 11 & 18 & 43\\
\sidehead{ISCS}
J1429.2$+$3357 & 14:29:15.16 & $+$33:57:08.5 & 1.059 & $\cdots$\tablenotemark{b} & $\cdots$ & $\cdots$ & $\cdots$ &  & 22 & 13 & 9 & 12 & 10\\
J1432.4$+$3332 & 14:32:29.18 & $+$33:32:36.0 & 1.112 & 4.9$_{-1.2}^{+1.6}$ & $734\pm115$ & 4 & 16 &  & 11 & 4 & 7 & 8 & 3\\
J1426.1$+$3403 & 14:26:09.51 & $+$34:03:41.1 & 1.136 & $\cdots$\tablenotemark{b} & $\cdots$ & $\cdots$ & $\cdots$ &  & 20 & 10 & 10 & 10 & 10\\
J1426.5$+$3339 & 14:26:30.42 & $+$33:39:33.2 & 1.163 & $\cdots$\tablenotemark{b} & $\cdots$ & $\cdots$ & $\cdots$ &  & 28 & 15 & 13 & 19 & 9\\
J1434.5$+$3427 & 14:34:30.44 & $+$34:27:12.3 & 1.238 & 2.5$_{-1.1}^{+2.2}$ & $863\pm170$ & 4 & 4 &  & 21 & 10 & 11 & 13 & 8\\
J1429.3$+$3437 & 14:29:18.51 & $+$34:37:25.8 & 1.262 & 5.4$_{-1.6}^{+2.4}$ & $767\pm295$ & 4 & 4 &  & 23 & 15 & 8 & 16 & 7\\
J1432.6$+$3436 & 14:32:38.38 & $+$34:36:49.0 & 1.349 & 5.3$_{-1.7}^{+2.6}$ & $807\pm340$ & 4 & 4 &  & 21 & 8 & 13 & 14 & 7\\
J1433.8$+$3325 & 14:33:51.13 & $+$33:25:51.1 & 1.369 & $\cdots$\tablenotemark{b} & $\cdots$ & $\cdots$ & $\cdots$ &  & 19 & 8 & 11 & 17 & 2\\
J1434.7$+$3519 & 14:34:46.33 & $+$35:19:33.5 & 1.372 & 2.8$_{-1.4}^{+2.9}$ & $\cdots$ & 4 & $\cdots$ &  & 18 & 5 & 13 & 16 & 2\\
J1438.1$+$3414 & 14:38:08.71 & $+$34:14:19.2 & 1.414 & 3.1$_{-1.4}^{+2.6}$ & $757_{-208}^{+247}$ & 4 & 5 &  & 38 & 15 & 23 & 32 & 6\\
J1432.4$+$3250 & 14:32:24.16 & $+$32:50:03.7 & 1.487 & 2.5$_{-0.9}^{+1.5}$ & $\cdots$ & 5 & $\cdots$ &  & 31 & 12 & 19 & 25 & 6\\
\cline{1-14}
Total Number &  &  &  &  &  &  &  &  & 1386 & 1075 & 311 & 392 & 994
\enddata
\tablenotetext{a}{A published value of $M_{200}$ for MACS 2129.4$-$0741 could not be found.}
\tablenotetext{b}{ISCS clusters without published $M_{200}$ values.}
\tablerefs{(1) \cite{umetsu2016}; (2) \cite{lemze2013}; (3) \cite{merten2015};
(4) \cite{jee2011}; (5) \cite{brodwin2011}; (6) \cite{geller2014};
(7) \cite{annunziatella2016}; (8) \cite{rines2013}; (9) \cite{gomez2012};
(10) \cite{balestra2016}; (11) \cite{biviano2013}; (12) \cite{lu2010};
(13) \cite{ebeling2007}; (14) \cite{ma2008}; (15) \cite{jorgensen2013};
(16) \cite{eisenhardt2008}.}
\end{deluxetable*}

For our $0.15<z<1$ cluster sample, we use the publicly available\footnote{\url{https://archive.stsci.edu/prepds/clash/}}
CLASH survey \citep{postman2012}, which provides observations of
25 clusters in 16 \textit{Hubble Space Telescope} (\hst) bands spanning
the ultraviolet to the near-infrared, and four \spitzer/IRAC near-infrared
bands. While the main scientific goal of CLASH was to constrain the
mass distributions of galaxy clusters using their gravitational lensing
properties \citep[e.g.,][]{merten2015}, a number of  ancillary works
have taken advantage of the wealth of multi-wavelength coverage, including
the study of the morphologies and SFRs of brightest cluster galaxies
\citep[BCGs;][]{donahue2015}, the investigation of galaxies potentially
undergoing ram-pressure stripping \citep{mcpartland2016}, and the
characterization of high-redshift core-collapse supernovae rates \citep{strolger2015}.
Our use of the high-quality \hst\ and \spitzer/IRAC observations
to derive physical properties (SFRs and stellar masses) enables us
to study the evolution of SF activity in CLASH galaxies.

The high-redshift ($1<z<1.5$) portion of our cluster sample is based
on the ISCS \citep{eisenhardt2008}, which identified $0.1<z<2$ galaxy
clusters in 7.25 deg$^{2}$ of the Bo\"otes field of the NOAO Deep
Wide-Field Survey \citep[NDWFS;][]{jannuzi1999}. A wavelet algorithm
was used, with accurate photometric redshifts from \citet{brodwin2006},
to isolate three-dimensional overdensities of $4.5\,\mu$m-selected
galaxies, with the cluster centers being identified as the peaks in
the wavelet detection maps. We use 11 spectroscopically confirmed
clusters over $1<z<1.5$, all of which have deep \hst\ observations
\citep{snyder2012} and were studied in Paper I. For a more detailed
description of our high-redshift cluster sample, please consult Paper
I.

\subsection{Cluster Halo Mass}

We now describe cluster halo masses, $M_{200}$, compiled from the
literature, and briefly address how ISCS clusters are likely progenitors
of CLASH-mass halos. $M_{200}$ is the mass contained in a radius
$R_{200}$, which defines a region where the density is 200 times
the critical density of the Universe, $\rho_{\mathrm{crit}}$.

\citet{umetsu2016} measured $M_{200}$ values for 20 of the 25 CLASH
clusters using a combined strong-lensing, weak-lensing, and magnification
analysis. We take the $M_{200}$ measurement of A1423 from \citet{lemze2013},
who used the projected mass profile estimator, a dynamical method,
to derive the cluster mass. Using a combined weak- and strong-lensing
analysis, \citet{merten2015} measured $M_{200}$ for 19 CLASH systems.
We use their estimate of $M_{200}$ for MACS1311, MACS1423, and CLJ1226.
For MACS2129, a non-exhaustive search of the literature did not reveal
a published $M_{200}$.\footnote{\citet{mantz2010} report an $M_{500}$ of $\sim$$10^{15}\ M_{\odot}$,
based on gas mass derived from X-ray observations.}

\begin{figure}[!t]
\includegraphics{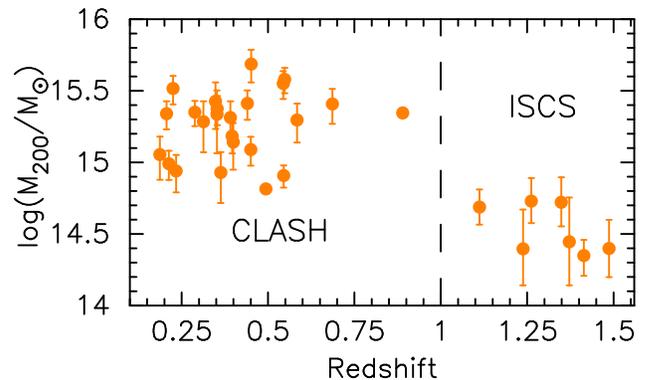}

\caption{Cluster $M_{200}$ versus redshift for the 31 clusters for which we
have halo mass estimates from the literature. Error bars on $M_{200}$
values are taken from the literature. The dashed line separates the
CLASH and ISCS subsets.\label{fig:z_M200}}
\end{figure}

$M_{200}$ estimates for six of our ISCS clusters are taken from \citet{jee2011},
who used a weak-lensing analysis. \citet{brodwin2011} measured the
cluster mass of ISCS J1432.4$+$3250 using \chandra\ X-ray observations. For the remaining ISCS clusters in our sample, which do not have published M$_{200}$ estimates, their halo masses can be taken to be the average sample halo mass of $M_{200}\sim10^{14}\,M_{\odot}$, inferred from an analysis of the correlation function of the entire set of ISCS clusters \citep{brodwin2007}.

Figure~\ref{fig:z_M200} shows $M_{200}$ as a function of redshift
for each cluster in our sample, excluding those for which we do not
have published mass estimates. Given the large differences between
the $M_{200}$ of CLASH and ISCS clusters, we must now determine whether
the latter clusters could be the likely progenitors of any of the
24 CLASH clusters for which we have $M_{200}$ estimates. Should that
be the case, it would follow that the galaxies that reside in these
high-redshift clusters can be considered likely progenitors of the
CLASH galaxies at $z<1$.

We model the mass growth of our clusters across the entire redshift
range of our sample with the code COncentration-Mass relation and
Mass Accretion History \citep[COMMAH;][]{correa2015a,correa2015b,correa2015c},
which uses an analytic model to generate halo mass accretion rates
for a variety of redshifts ($\Delta z=0.25$) and cluster masses ($\Delta\log\left(M_{200}\right)=0.2$).
In addition to the code\footnote{\url{http://ph.unimelb.edu.au/~correac/html/codes.html} or \url{https://github.com/astroduff/commah}.}
itself, \citet{correa2015a,correa2015b,correa2015c} provide tabulated
accretion rates for a number of different cosmologies.\footnote{While the $d\left(M_{200}\right)/dt$ we use were tabulated for a
WMAP9 cosmology \citep{hinshaw2013}, the differences in cluster masses
are negligible between this and WMAP7.}

\begin{deluxetable}{cccc}
\tablecaption{$M_{200}$ Evolutionary Track Parameters\label{tab:clusterEvolutionaryTracks}}
\tablewidth{0pt}
\tablehead{
\colhead{Cluster} &
\colhead{$t_0$} &
\colhead{$M_{200}$} &
\colhead{$d\left(M_{200}\right)/dt_{\mathrm{L}}$}
\tabularnewline
\colhead{} &
\colhead{(Gyr)} &
\colhead{($10^{14}\,M_{\odot}$)} &
\colhead{$M_{\odot}\,\mathrm{yr}^{-1}$}
}
\startdata
MACS 1311.0$+$0310 & $5.0$ & $6.5$ & $10^{5.0}$\tabularnewline
MACS 0744.9$+$3927 & $6.3$ & $25.6$ & $10^{5.9}$\tabularnewline
ISCS J1429.3$+$3437 & $8.7$ & $5.4$ & $10^{5.5}$\tabularnewline
ISCS J1432.4$+$3250 & $9.4$ & $2.5$ & $10^{5.1}$\tabularnewline
\enddata
\end{deluxetable}

We compare our sample to their tabulated grids, selecting accretion
rates for four clusters where both the total masses and redshifts
agree well (differences of no more than 0.07 and 0.06 in $\log\left(M_{200}/M_{\odot}\right)$
and redshift, respectively). We list these clusters in Table\ \ref{tab:clusterEvolutionaryTracks},
along with the lookback time at their redshift, $M_{200}$ at their
redshift, and mass accretion rate. With the given accretion rate, a cluster's mass can
be projected as a function of time through

\begin{equation}
M_{200}\left(t_{\mathrm{L}}\right)=-\left(t_{\mathrm{L}}-t_{0}\right)\frac{d\left(M_{200}\right)}{dt_{\mathrm{L}}}+M_{200}\left(t_{0}\right),\label{eq:evolutionaryTrackEq}
\end{equation}
where $t_{\mathrm{L}}$ is the lookback time (the negative slope comes
from this dependence), $d\left(M_{200}\right)/dt_{\mathrm{L}}$ is
the mass accretion rate, $M_{200}\left(t_{0}\right)$ is a cluster's
mass at its observed redshift, and $t_{0}$ is the lookback time at
that redshift.

\begin{figure}[!t]
\includegraphics{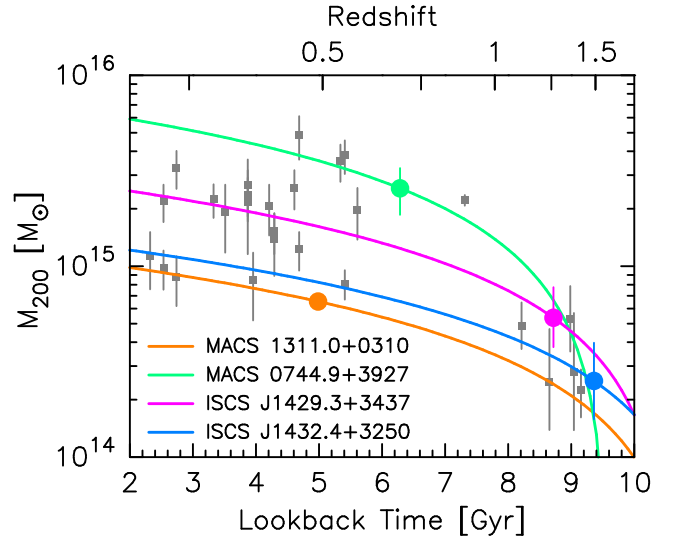}

\caption{Cluster $M_{200}$ versus lookback time for clusters in our sample
that have published mass estimates. The circles highlight the clusters
listed in Table\ \ref{tab:clusterEvolutionaryTracks}, with the corresponding
colored curves showing their projected mass evolution, described by
Equation\ (\ref{eq:evolutionaryTrackEq}) with parameters from Table\ \ref{tab:clusterEvolutionaryTracks}.\label{fig:clusterMassEvolution}}
\end{figure}

Figure\ \ref{fig:clusterMassEvolution} shows the evolutionary tracks
based on Equation\ (\ref{eq:evolutionaryTrackEq}) for our selected
clusters. Also shown are the measured cluster $M_{200}$ values as
a function of lookback time, with the bulk of our sample plotted with
gray squares. The four chosen clusters are shown with circles that
match the color of their respective time-dependent mass track.

Figure\ \ref{fig:clusterMassEvolution} can be understood in two
ways. It may either tell us about the unevolved mass of a cluster
at earlier lookback times considering its current (low-redshift) mass,
or it may inform us about the evolution of a high-redshift cluster's
$M_{200}$ up to recent times. Considering first the two selected
ISCS clusters, their tracks are found to intersect with a number of
CLASH clusters (at least within their uncertainties). For instance,
given its expected mass accretion, the cluster ISCS J1432.4$+$3250
at $z=1.49$ (blue curve) could conceivably evolve into an A383-sized
cluster ($M_{200}\sim10^{15}\,M_{\odot}$). Tracing the potential
evolution of ISCS J1429.3$+$3437 at $z=1.26$ (magenta curve), it
could also well evolve into a massive cluster, up to $\sim$$2\times10^{15}\,M_{\odot}$.

While neither of the ISCS tracks overlap with the most massive CLASH
clusters, we now consider the backwards evolution (with increasing
lookback time) of the $z=0.69$ cluster MACS0744 (green curve). Given
its projected growth rate, it could be considered as a mid-evolutionary
phase between the low-mass clusters we observe at $t_{\mathrm{L}}\sim9$
Gyr and the $\sim$$\left(3-5\right)\times10^{15}\,M_{\odot}$ CLASH
clusters. Based on its predicted evolution, MACS1311 ($z=0.49$; orange
curve) provides a link between the low-mass (for our samples) cluster
regime at all considered lookback times/redshifts. The good agreement
seen in Figure\ \ref{fig:clusterMassEvolution} between the measured
and inferred cluster $M_{200}$ values suggests that ISCS members
are, in an aggregate sense, the likely progenitors of CLASH cluster
galaxies.

\section{Galaxy Data and Sample Selection}

\label{sec:sample_selection}

We now describe the selection of cluster members (Section\ \ref{sub:cluster_membership}),
morphological classification of CLASH and ISCS galaxies (Section\ \ref{sub:Morphology}),
removal of BCGs (Section\ \ref{sub:BCGs}) and galaxies that host
AGNs (Section\ \ref{sub:agn}), and the measurement of SFRs and stellar
masses (Section\ \ref{sec:SFR_and_mass}).

\subsection{Selecting Cluster Members}

\label{sub:cluster_membership}

\subsubsection{ISCS}

In Paper I, the selection of ISCS cluster members used two different
methods. First, galaxies were considered members if they had high-quality
spectroscopic redshifts consistent with ISCS cluster redshifts \citep{eisenhardt2008,brodwin2013,zeimann2013}.
Second, if no high-quality $z_{\mathrm{spec}}$ was available, ISCS
galaxies were deemed members if they belong to the cluster red-sequence
as determined by \citet{snyder2012} with \hst\ photometry. Of the 270 ISCS cluster members, 69 are selected with spectroscopic redshifts, and 201 based on galaxy colors. Using the published table of spectroscopic redshifts from \citet{zeimann2013}, we identify 43 confirmed non-members that were included in the red-sequence analysis of \citet{snyder2012}. Of these galaxies, 11 would have been considered cluster members based on their color, for a field contamination rate of $\sim$26\%.

\subsubsection{CLASH}

\label{sub:CLASH_redshifts}

\begin{figure*}[!t]
\begin{centering}
\includegraphics{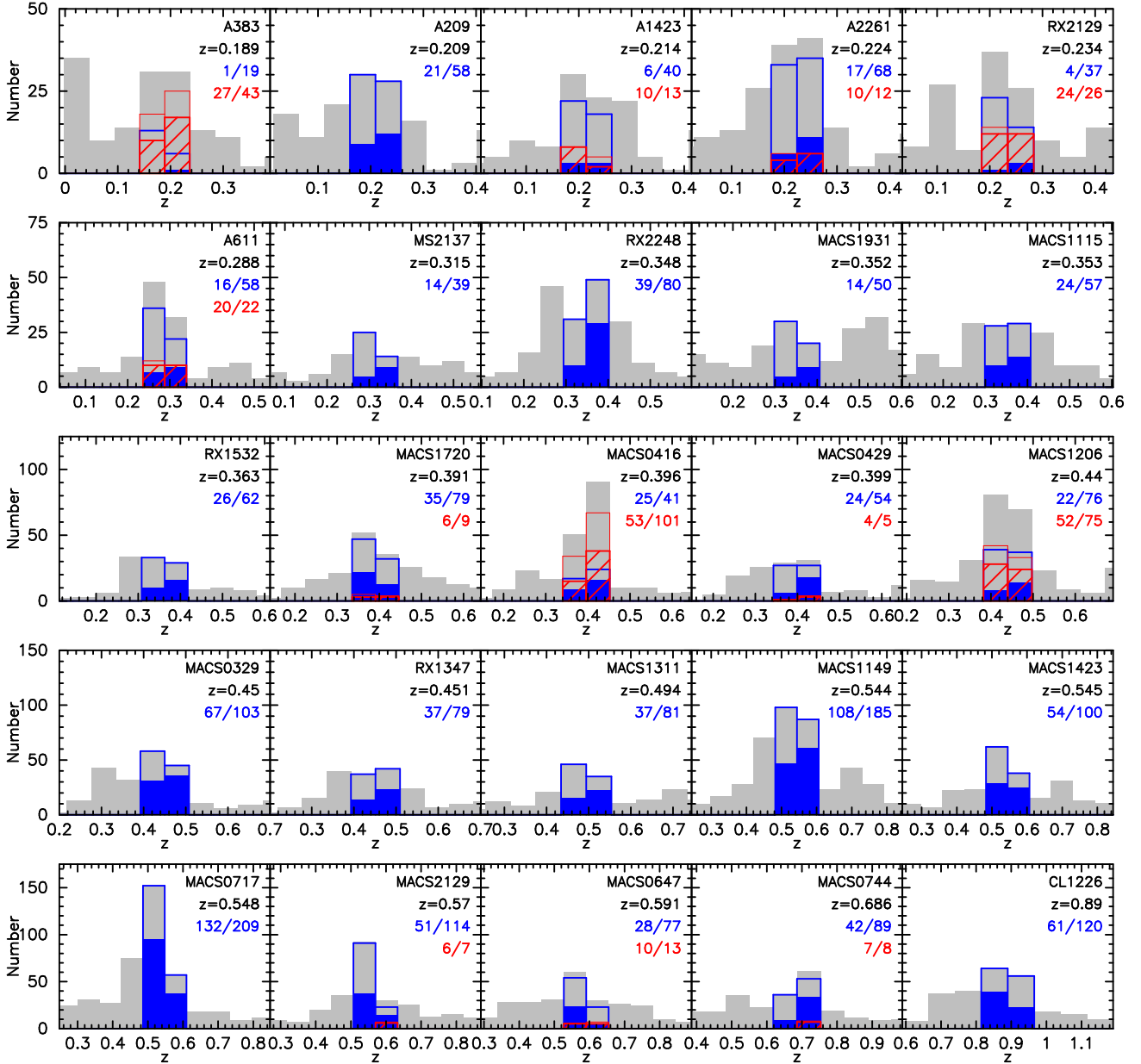}
\par\end{centering}

\caption{Histograms of $z_{\mathrm{spec}}$ (red open) and $z_{\mathrm{phot}}$
(blue open) CLASH members, superimposed over the best available redshift
of the entire CLASH sample (gray). Filled blue (hashed red) histograms
show the final sample of CLASH $z_{\mathrm{phot}}$ ($z_{\mathrm{spec}}$)
members. Each panel shows the shortened cluster name, $z_{\mathrm{cluster}}$,
and final number of $z_{\mathrm{phot}}$ (blue) and $z_{\mathrm{spec}}$
members (red) over the total number of each. Members not meeting the
cuts described throughout the text are not plotted.\label{fig:CLASH_zDist}}
\end{figure*}

As part of our CLASH membership selection, we prune likely stars from
our final sample by removing objects with the SExtractor \citep{bertin1996}
stellarity parameter $\mathrm{CLASS\_STAR}>0.85$.

\paragraph{Spectroscopic Redshifts}

While all CLASH clusters likely have spectroscopic redshifts measured
for a small number of members, the literature provides extensive catalogs
of publicly available spectroscopic redshift measurements for galaxies
in a subset of the 25 CLASH systems. The selections, described below,
result in 334 spectroscopic members.

The ongoing CLASH-VLT program \citep{rosati2014} will provide an
unprecedented number of spectroscopic redshifts in 24 of the 25 CLASH
fields once completed. We take advantage of spectroscopic redshifts
for the two CLASH-VLT clusters publicly released: MACS0416 \citep{balestra2016}
and MACS1206 \citep{biviano2013}.\footnote{Released CLASH-VLT spectroscopic redshift catalogs are available at
\url{https://sites.google.com/site/vltclashpublic/}.} For MACS1206, we use the list of cluster members provided by \citet{girardi2015}, where the membership was determined using
the technique for selecting cluster members from \citet{fadda1996}.
In this method, the peak in the spectroscopic redshift distribution
is found and galaxies are iteratively rejected if their redshifts
lie too far from the main distribution. \citet{balestra2016}, who
used the same method for determining cluster membership, did not include
a membership flag in the MACS0416 catalog. However, as all of our
potential MACS0416 members are contained within $R_{\text{proj}}\lesssim0.7\,\mathrm{Mpc}$,
we use $cz\leq\pm2100\,\mathrm{km}\,\mathrm{s}^{-1}$ as our membership
criterion, based on the extent of cluster members at that radius \citep[see Figure 6 of][]{balestra2016}.

We use spectroscopic redshifts for A383 from \citet{geller2014},
A611 from \citet{lemze2013}, and for A1423, A2261, and RXJ2129 from
\citet{rines2013}. For each of these five clusters, membership was
determined via caustic techniques \citep{diaferio97,diaferio99}.
For A383, A1423, A2261, and RXJ2129, the authors included a membership
flag in their published redshift tables; we use this flag to select
spectroscopic members for these four clusters. For A611, spectroscopic
members are based on an unpublished membership list (D. Lemze, private
comm.).

We further supplement our catalog with spectroscopic redshifts for
candidate members in MACS1720, MACS0429, MACS2129, MACS0647, and MACS0744
from the catalog published by \citet{stern2010}. After removing star-like
objects, we include as cluster members galaxies with $-0.02<z_{\mathrm{spec}}-z_{\mathrm{cluster}}<0.03$,
based on the peak in the distribution of redshift residuals.

\paragraph{Photometric Redshifts}

To select the remainder our our $z<1$ cluster members we use photometric
redshift estimates derived by \citet{postman2012} using Bayesian
Photometric Redshift \citep{benitez2000,benitez2004,coe2006}, a Bayesian
$\chi^{2}$ minimization template fitting software package. Each galaxy's
photometric redshift was given an odds parameter, $P\left(z\right)$,
defined as the integral of the probability distribution in a region
of $\sim$$2\left(1+z_{\mathrm{phot}}\right)$ around $z_{\mathrm{phot}}$.
A value near unity indicates that the distribution has a narrow single
peak \citep{benitez2004} and our final sample only contains galaxies
with $P\left(z\right)>0.9$, considered to be the most reliable by
\citet{postman2012}. For a galaxy to be a cluster member, we require
that its $z_{\mathrm{phot}}$ be contained within $z_{\mathrm{cluster}}\pm0.04\times\left(1+z_{\mathrm{cluster}}\right)$.

In order to estimate the potential field contamination of selecting cluster members by their photometric redshifts we identify the 172 galaxies in our initial CLASH sample that are confirmed non-members according to their spectroscopic redshifts. Of these, 45 have $z_{\mathrm{phot}}$ consistent with being members, for a field contamination rate of $\sim$26\%.

\paragraph{Cluster Members}

Figure\ \ref{fig:CLASH_zDist} shows the distribution of photometric
and spectroscopic redshifts for the CLASH members. The 334 $z_{\mathrm{spec}}$
members are plotted as the red open histograms, while the 1975 $z_{\mathrm{phot}}$
members are depicted by the blue open histograms. Cluster names and
redshifts are listed. Due to cuts on BCGs (Section\ \ref{sub:BCGs}),
presence of an AGN (Section\ \ref{sub:agn}), morphology (Section\ \ref{sub:Morphology}),
filter requirements for SED fitting (Section\ \ref{sub:SED_fitting}),
and stellar mass (Section\ \ref{sub:SFR_and_mass}) the final CLASH
sample contains 1134 galaxies. The red hatched (blue filled) histograms
show the 229 (905) final spectroscopic (photometric) members.

\subsection{Galaxy Morphology}

\label{sub:Morphology}

\begin{figure*}[!t]
\begin{centering}
\includegraphics{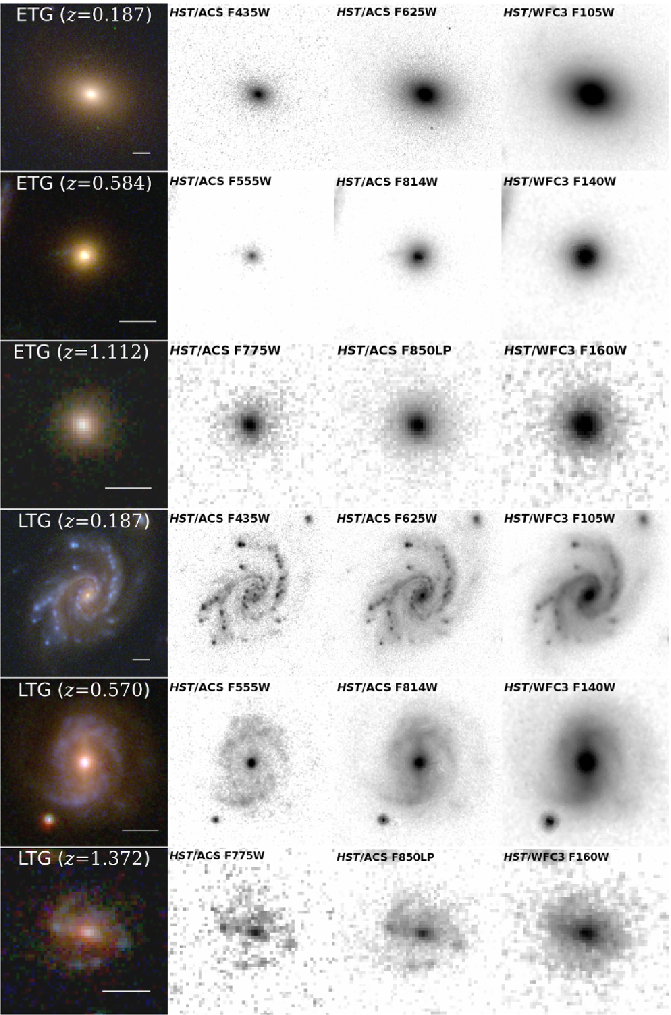}
\par\end{centering}

\caption{Pseudocolor cutouts (left column) of example visually classified ETGs
(top three rows) and LTGs (bottom three rows), with morphology and
cluster redshift listed for each galaxy. A scale of $1\arcsec$ is
shown by the white line in each pseudocolor image. The right three
columns show the blue, green, and red filters that comprise each pseudocolor
image and that were used for visual classification. Each image is
30x30 kpc.\label{fig:ETGmorphology}}
\end{figure*}

In Paper I, galaxies were visually classified using observed-frame
\hst\ optical images, taken with either the Advanced Camera for Surveys
\citep[ACS;][]{ford1998} or the Wide Field Planetary Camera 2 \citep{holtzman1995},
and near-infrared images, acquired with the Wide Field Camera 3 \citep[WFC3;][]{kimble2008}.
ETGs were required to have a smooth elliptical/S0 shape, while galaxies
that showed clear spiral structure, or irregular or disturbed morphologies,
were collectively classified as LTGs. For consistency with Paper I,
we similarly visually classify the morphology of CLASH members. Given
the broad spectral coverage afforded by CLASH, we simultaneously inspect
at least one \hst\ image in each of the ultraviolet, optical and
near-infrared. We could not cleanly assess the morphology for a small
subset of galaxies. The latter are thus left out of any forthcoming
analysis.

Figure\ \ref{fig:ETGmorphology} shows examples of visually classified
ETGs and LTGs, spanning the majority of the redshift range of our
sample. While the visual classification was performed by simultaneously
inspecting single-band \hst\ images (shown in the three rightmost
columns, with filter listed), we also provide pseudocolor cutouts
in the left column.

\newpage
\subsection{Excluding Brightest Cluster Galaxies}

\label{sub:BCGs}

BCGs are typically the dominant, central galaxy in the cluster environment
and are thought to have undergone an evolution different to that of
the rest of the cluster population \citep[and references therein]{groenwald2014,donahue2015}.
Because of their potentially different histories and sometimes elevated
SF activity \citep[e.g.,][]{mcnamara2006,mcdonald2013,mcdonald2016},
we remove them from our analysis.

The positions of all galaxies in our CLASH sample are compared with
the CLASH BCGs studied by \citet[their Table 4]{donahue2015}. We
do not find a match for eight of their BCGs as our own selection criteria
already singled out these galaxies. We inspect each pair of the remaining
matched sources in rest-frame ultraviolet and optical \hst\ images
to ensure that the object is indeed the cluster BCG. This results
in the elimination of 17 additional BCGs across all 25 CLASH clusters.

In a manner similar to the BCG identification method of \citet{lin2013}
we identify potential ISCS BCGs by first selecting the $\sim$5 galaxies in each of the 11 ISCS cluster fields that have the highest $4.5\,\mu$m flux. However, none of these 56 galaxies match our final list of ISCS members. We are confident that our cluster sample
is free of BCGs.

\subsection{AGN Removal}

\label{sub:agn}

Since AGNs contribute to galaxies' infrared flux and can potentially
contaminate our SFRs and stellar masses, we remove their host galaxies
from our sample. Following Paper I, we identify AGNs using the IRAC
color-color selection from \citet{stern2005}. Unlike that treatment,
however, we do not remove galaxies with fewer than four good IRAC
fluxes. Additionally, all ISCS galaxies with hard X-ray luminosities
exceeding $10^{43}\,\mathrm{erg}\,\mathrm{s}^{-1}$ are removed. This results in the removal of 17 cluster members.

\subsection{Estimating Physical Properties of Cluster Galaxies}

\label{sec:SFR_and_mass}

\subsubsection{ISCS}

Paper I used IRAC observations as positional priors to match the MIPS
fluxes to \hst\ catalog sources. The \citet{chary2001} templates
were used to convert the measured 24 $\mu$m fluxes to total IR luminosities,
then SFRs were calculated using relation\ (4) from \citet{murphy2011}.
Using \hst\ images, ISCS members were visually inspected to determine
isolation. Galaxies with nearby neighbors were removed from the final
sample, as their mid-infrared SFRs could not be reliably measured
due to the large MIPS point spread function. For consistency with
our CLASH sample, here we calculate ISCS SFRs through SED fitting.
We describe this procedure in Section\ \ref{sub:SED_fitting}, and
in Appendix\ \ref{sec:SFR_mass_comparison} we present a comparison
of these SFRs and the 24 $\mu$m SFRs from Paper I.

Stellar masses were estimated in Paper I by fitting galaxies' SEDs
to population synthesis models using \texttt{iSEDfit} \citep{moustakas2013},
the \citet{bruzual2003} stellar population models, and the \citet{chabrier2003}
IMF. As with the ISCS SFRs, we re-derive stellar masses as described
in the following section, and compare them to the values calculated
with \texttt{iSEDfit} in Appendix\ \ref{sec:SFR_mass_comparison}.

Photometry for our ISCS galaxies includes three NDWFS optical wavebands
($B_{w}$, $R$, and $I$), three near-infrared bands from the FLAMINGOS
Extragalactic Survey \citep[$J,\;H$, and $K_s$;][]{elston2006}, one
near-infrared \hst\ band (F160W), and five \spitzer\ bands (3.6,
4.5, 5.8, and 8.0 $\mu$m on IRAC, and 24 $\mu$m on MIPS). Despite
the small number of bands relative to CLASH, the ultraviolet portion
of the spectrum is well-covered at $z\sim1$ by the $B_{w}$ and $I$
filters, with $R$ band straddling the ultraviolet and optical regimes.
F160W and all four bands of IRAC provide good coverage of the infrared.
In the range of $z\sim1.5$, F814W and $R$ now also sample the ultraviolet,
while F160W falls into the redshifted optical regime. While we do
have 24 $\mu$m MIPS observations, we exclude them from our final
SED fits in order to be more inclusive of non-isolated galaxies. We
discuss the validity of this choice in Appendix\ \ref{sec:SFR_mass_comparison}.

\subsubsection{Spectral Energy Distribution Fitting}

\label{sub:SED_fitting}

To estimate the SFRs and stellar masses for all galaxies in our sample
we use the Code Investigating GALaxy Emission \citep[CIGALE;][]{burgarella2005,noll2009,roehlly2012,roehlly2014}.\footnote{CIGALE is publicly available at \url{http://cigale.oamp.fr/}.}
Combining various models, each encompassing a number of free parameters,
CIGALE builds theoretical SEDs, using the energy balance approach,
in which (conceptually) a portion of a galaxy's flux is absorbed in
the ultraviolet with a corresponding re-radiation in the infrared,
with the latter balancing the total energy of the system. CIGALE then
compares the theoretical SEDs to multi-wavelength observations, generating
a probability distribution function for each parameter of interest.

The input data for each galaxy in our sample are its unique identifier,
redshift, and the flux and associated errors in each observed filter.
With CIGALE's energy balance approach in mind, we only include galaxies
that have good fluxes in at least one rest-frame filter in each the
ultraviolet and near-infrared. To reduce computational time during
the SED modeling, we set the redshift of each member galaxy to its
parent cluster's redshift. For all CIGALE runs we adopt a \citet{bruzual2003}
stellar population model, with a \citet{chabrier2003} IMF, and solar
metallicity. Table\ \ref{tab:CIGALEparams} lists the CIGALE parameters
that define the template SEDs for all galaxies.

\begin{deluxetable}{cc}
\tablecaption{CIGALE Input Parameters\label{tab:CIGALEparams}}
\tablewidth{0pt}
\tablehead{
\colhead{Parameter} &
\colhead{Values}
}
\startdata
\multicolumn{2}{c}{Double declining exponential star formation history} \\
\cline{1-2}
$\tau_{\mathrm{main}}$ (Gyr) & $0.5-5$ \\
$\tau_{\mathrm{burst}}$ (Gyr) & 10 \\
$t_{\mathrm{main}}$ (Gyr)\tablenotemark{a} & $1-11$ \\
$t_{\mathrm{burst}}$ (Myr) & $50-750$ \\
$f_{\mathrm{burst}}$ & $0-0.2$ \\
\cline{1-2}
\multicolumn{2}{c}{Dust attenuation: \citet{calzetti2000} and \citet{leitherer2002}} \\
\cline{1-2}
$E\left(B-V\right)_{\star,\mathrm{young}}$\tablenotemark{b} & $0-1$ \\
\cline{1-2}
\multicolumn{2}{c}{Dust template: \citet{dale2014}} \\
\cline{1-2}
$f_{\mathrm{AGN}}$ & 0 \\
$\alpha_{\mathrm{dust}}$ & $1-3$ \\
\enddata
\tablenotetext{a}{At a given redshift $t_{\mathrm{main}}$ never exceeds the age of the Universe.}
\tablenotetext{b}{Following \citet{buat2014}, a 50\% reduction factor is applied to the $E\left(B-V\right)_{\star}$ for old stellar populations ($\mathrm{age}>10\,\mathrm{Myr}$).}
\end{deluxetable}

Star formation histories (SFHs) commonly found in the literature include
a single stellar population with an exponentially declining SFR, a
single stellar population with an SFR that declines exponentially
after some delay time, or a two-component exponentially declining
model consisting of a recently formed young stellar population on
top of an underlying old stellar population. As galaxies are expected
to undergo multiple SF episodes, this latter type of SFH is better
suited than a single-population SFH to model complex stellar systems
\citep{buat2014}. However, \citet{ciesla2016} have recently shown
promising results using a \textit{truncated} delayed SFH to model
the rapid quenching expected of galaxies undergoing ram-pressure stripping
in the cluster environment. While a comparison of the physical properties
of CLASH galaxies derived by different SFHs will be reviewed elsewhere,
we adopt a double declining exponential SFH in this work. 

The SFR in a declining exponential SFH is characterized by

\begin{equation}
SFR\left(t\right)\propto e^{-t/\tau},\label{eq:1tau}
\end{equation}
where $t$ is time (i.e., the age of the stellar population) and $\tau$
is the $e$-folding time. As our adopted SFH comprises an old (main)
and younger (burst) stellar population, each has an age ($t_{\mathrm{main}}$
and $t_{\mathrm{burst}}$, respectively) and an $e$-folding time
($\tau_{\mathrm{main}}$ and $\tau_{\mathrm{burst}}$, respectively).
Prior to the burst occurring ($t<t_{\mathrm{main}}-t_{\mathrm{burst}}$),
the galaxy forms stars according to Equation\ (\ref{eq:1tau}). Subsequently,
the SFR of the galaxy is modeled as a linear combination of two such
functions, with the exponential component for the young stellar population
modulated by a burst fraction ($f_{\mathrm{young}}$), which defines
the mass fraction contributed by young stars. To generate a wide array
of possible galaxy histories, we supply CIGALE with a range of values
for $t_{\mathrm{burst}}$, $t_{\mathrm{main}}$, and $\tau_{\mathrm{main}}$.
We approximate a constant burst of SF for the young population by
fixing its $e$-folding time at 10 Gyr. The remaining parameters listed
in Table\ \ref{tab:CIGALEparams} are the dust attenuation of the
young population ($E(B-V)_{\star,\mathrm{young}}$), the fraction
of infrared emission due to an AGN ($f_{\mathrm{AGN}}$), and the
infrared power law slope ($\alpha_{\mathrm{dust}}$). As we remove
AGNs from our entire cluster sample (Section\ \ref{sub:agn}), we
fix $f_{\mathrm{AGN}}=0$ for all models. While we have removed galaxies with a dominating AGN component, there may be some subdominant AGN contribution to the measured SFRs of the remainder of our sample. However, disentangling such low-level contribution is beyond the scope of this work.

\subsubsection{Star Formation Rates and Stellar Masses}

\label{sub:SFR_and_mass}

\begin{figure}[!t]
\includegraphics{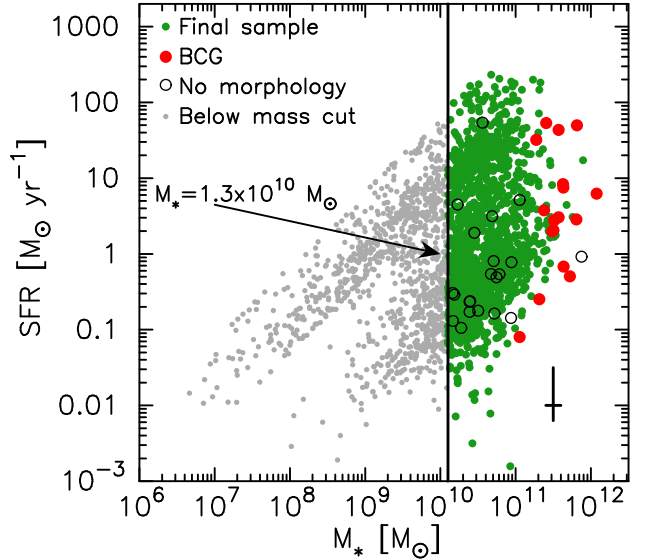}

\caption{CIGALE-derived SFR versus stellar mass for cluster members, with our
final sample plotted with green points. Also shown are galaxies cut
from the sample (see legend), as well as the stellar mass cut (vertical
line). Galaxies identified as likely AGN hosts (Section\ \ref{sub:agn})
are not plotted. The error bar in the lower right corner shows the typical uncertainty in the SFRs and stellar masses derived with CIGALE.\label{fig:z_mass_SFR_final}}
\end{figure}

Figure\ \ref{fig:z_mass_SFR_final} shows the CIGALE-derived SFRs
and stellar masses for our cluster sample. In addition to the sample
cuts described up to this point, we impose a further cut on our sample
by requiring all galaxies to meet the 80\% mass completeness limit
of Paper I, $\log\left(M_{\star}/M_{\odot}\right)>10.1$.\footnote{Because galaxies build up stellar mass over time, this constant cut
may introduce some bias in our sample selection. For example, the
progenitor of a $\log\left(M_{\star}/M_{\odot}\right)\sim10.1$ galaxy
at low redshift will likely have had a lower stellar mass at higher
redshift and hence be excluded from our sample. However, while the
ideal solution would involve an evolving stellar mass cut, our sample
size precludes adopting such a selection.} Galaxies that fall above our stellar mass cut (to the right of the
vertical line) for which we could not reliably determine a morphology
(Section\ \ref{sub:Morphology}) are plotted with open black circles.
We do not identify \textit{all} galaxies without a morphology as many
of the galaxies that fall below our mass cut were not inspected for
morphology. BCGs identified in Section\ \ref{sub:BCGs} are shown
by the red points and our final sample ($N=1386$) is shown by the
green circles. Table\ \ref{tab:clusterList} gives the breakdown
by cluster of the number of cluster members ($N_{\mathrm{tot}}$),
ETGs ($N_{\mathrm{ETG}}$) and LTGs ($N_{\mathrm{LTG}}$) in our final
sample.

\subsubsection{Separating Star-forming and Quiescent Galaxies}

\label{sub:SF_ and_quiescent_galaxies}

\begin{figure}[!t]
\includegraphics{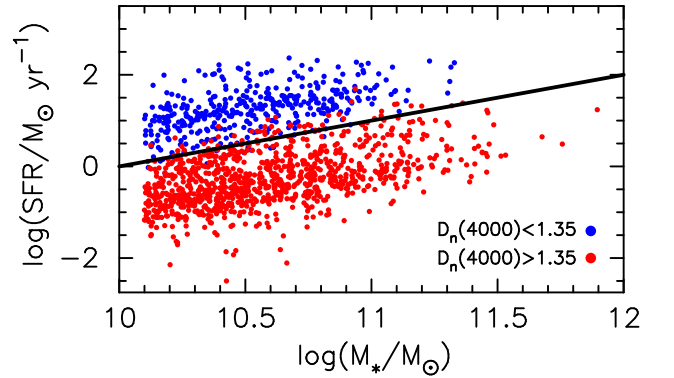}

\caption{SFR-$M_{\star}$ plane for SFGs (blue) and quiescent galaxies (red)
defined by a $D_{n}\left(4000\right)=1.35$ cut. The black line shows
a constant specific SFR of $SFR/M_{\star}=0.1\,\mathrm{Gyr}^{-1}$.\label{fig:SF_Q_testPlots}}
\end{figure}

In addition to classifying cluster galaxies based on their morphologies,
we separate members based on whether they are still actively forming
new stars (star-forming galaxies; SFGs) or are quiescent (i.e., have
little to no ongoing SF). These two galaxy types are commonly separated
on the basis of the strength of the 4000 \AA\ break, which is an
indicator of their stellar population age \citep{bruzual1983,balogh1999,kauffmann2003}.
The amplitude of the break, $D_{n}\left(4000\right)$, is used to
separate galaxies into young (star-forming) and old (quiescent) galaxies
\citep[e.g.,][]{vergani2008,woods2010}. Our SED fits enable a recovery
of $D_{n}\left(4000\right)$, which CIGALE calculates as the ratio
of the flux in the red continuum (4000--4100 \AA) to that in the
blue continuum (3850--3950 \AA), based on the definition of \citet{balogh1999}.

We adopt a cutoff of $D_{n}\left(4000\right)<1.35$ \citep{johnston2015}
to select SFGs, which results in 392 star-forming and 994 quiescent
cluster members. Table\ \ref{tab:clusterList} lists the final number
of SFGs ($N_{\mathrm{SFG}}$) and quiescent galaxies ($N_{\mathrm{Q}}$)
in each cluster, and Figure\ \ref{fig:SF_Q_testPlots} shows the
SFR versus stellar mass of each subset. Another common metric for
selecting SFGs \citep[e.g.,][]{lin2014} is the specific SFR ($sSFR\equiv SFR/M_{\star}$),
which measures a galaxy's efficiency at forming new stars. For comparison,
we find that a constant sSFR of $0.1\,\mathrm{Gyr}^{-1}$ (black line)
qualitatively agrees with our $D_{n}\left(4000\right)$ cut. We reinforce
that our selection of ETGs and LTGs in Section\ \ref{sub:Morphology}
was based on their \textit{morphologies} and should not to be confused
with spectroscopically early- and late-type galaxies, terms that are
sometimes used to refer to quiescent and star-forming galaxies, respectively
\citep[e.g.,][]{vergani2008}.

\section{Results and Discussion}

\label{sec:results}

Throughout this section, we incorporate into our error bars the uncertainty due to the potential field contamination of the cluster members selected with photometric redshifts or through red-sequence color selection (see Section\ \ref{sub:cluster_membership}). Error bars on fractional values are calculated as the quadrature sum of uncertainties based on binomial confidence intervals \citep{gehrels1986} and those due to potential field contamination. The latter are the differences between the calculated fraction and the bounds of the fraction assuming 26\% of the $z_{\mathrm{phot}}$/color-selected subset contained interlopers. The error bars on sSFR values are calculated as the quadrature sum of simple Poisson errors and 5000 iterations of bootstrap resampling. To calculate the former, the number of galaxies used in each bin is the sum of the number of $z_{\mathrm{spec}}$ members and 74\% of the number of $z_{\mathrm{phot}}$/color-selected members (to simulate a 26\% interloper rate). For each iteration of bootstrap resampling, pairs of SFR and stellar mass are chosen from the set of spectroscopically-selected members and a random sampling of 74\% of the non-$z_{\mathrm{spec}}$ members.

\subsection{The Relation Between SFR and Stellar Mass in Cluster SFGs}

\label{sub:SFR-mass_relation}

\begin{figure*}[!t]
\begin{centering}
\includegraphics{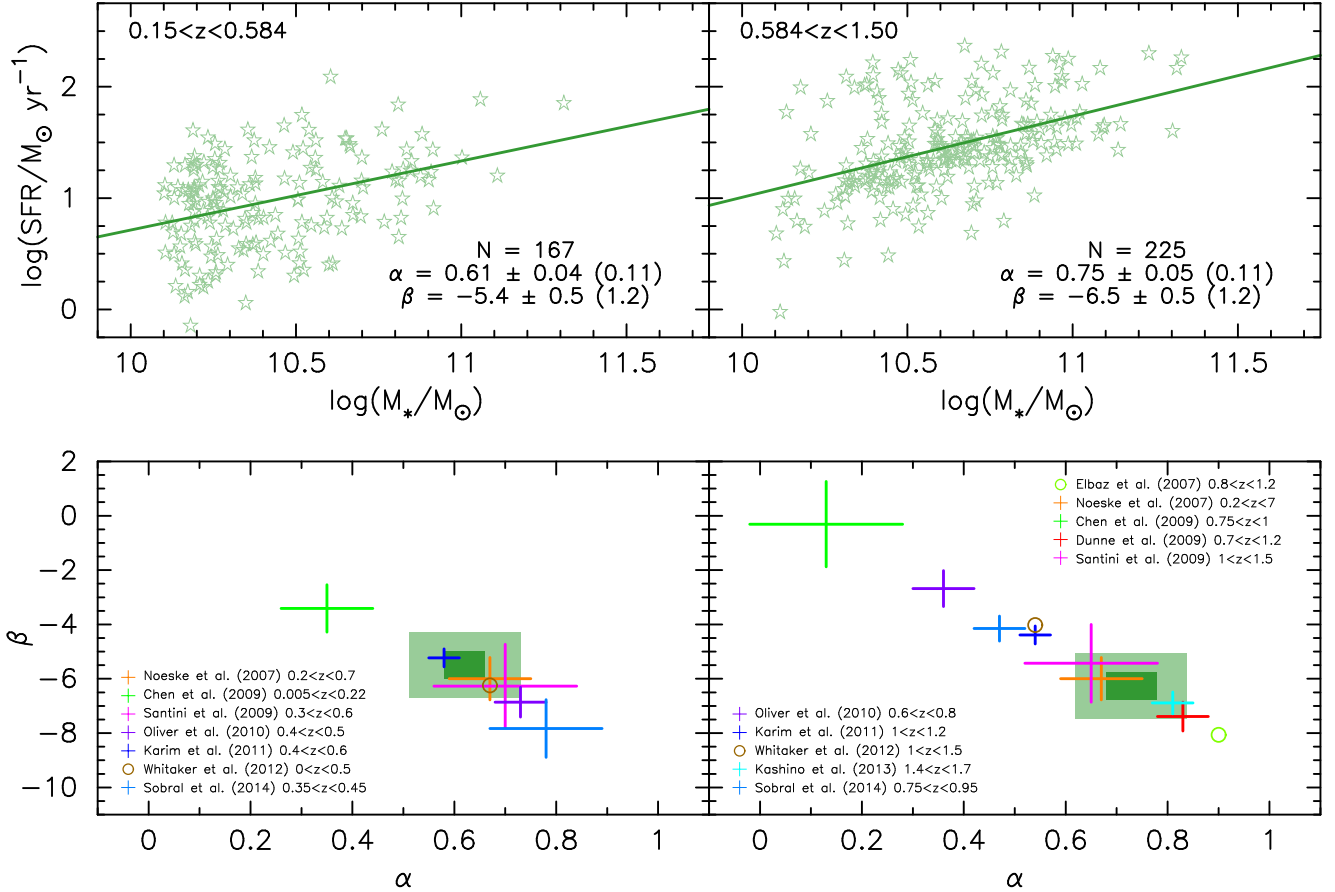}
\par\end{centering}

\caption{Upper panels: SFR versus $M_{\star}$ for cluster SFGs (light green stars). SFGs are fit (dark green lines) by Equation\ (\ref{eq:SFR_M_powerLaw}), with the number of galaxies, fit coefficients, and 1$\sigma$ and total (when accounting for potential field contamination; see text) uncertainties listed. Lower panels: Comparison of SFR-$M_{\star}$ parameters for our best fit, with 1$\sigma$ (dark green rectangles) and total (light green rectangles) uncertainties shown. A selection of literature best-fit parameters, calibrated by \citet{speagle2014}, are provided: \citet{elbaz2007}; \citet{noeske2007}; \citet{chen2009}; \citet{dunne2009}; \citet{santini2009}; \citet{oliver2010}; \citet{karim2011}; \citet{whitaker2012}; \citet{kashino2013}; and \citet{sobral2014}. Colored error bars show the parameter uncertainties (if available) from the literature; best-fit parameters with no uncertainties are shown by open circles.\label{fig:mass_SFR}}
\end{figure*}

SFR (or sSFR) as a function of stellar mass, hereafter referred to
as the SFR-$M_{\star}$ relation, has been studied extensively over
the last decade, resulting in a tight correlation between these two
physical quantities, often referred to as the ``main sequence''
of SF \citep{noeske2007,elbaz2011}. This is often characterized by
a power law of the form
\begin{equation}
\log\left(SFR/M_{\odot}\,\mathrm{yr}^{-1}\right)=\alpha\log\left(M_{\star}/M_{\odot}\right)+\beta,\label{eq:SFR_M_powerLaw}
\end{equation}
where $\alpha$ and $\beta$ are the slope and normalization, respectively.
The SFR-$M_{\star}$ relation as a function of redshift has been investigated
in numerous studies \citep[e.g.,][]{vulcani2010,ilbert2015,schreiber2015,tasca2015}.
\citet{speagle2014} compiled and standardized a sample of 25 studies
from the literature over the range of $0<z<6$ and found that despite
a variety of SFR and stellar mass indicators, there is a good consensus
on the SFR-$M_{\star}$ relations, with a 1$\sigma$ scatter of $\sim$0.1
dex among publications. Although \citet{speagle2014} calibrated the
literature fits to a \citet{kroupa2001} IMF, they note that differences
between this and the \citet{chabrier2003} IMF we employ have a negligible
impact on the coefficients of the SFR-$M_{\star}$ relation.

While studies of the SFR-$M_{\star}$ relation have largely focused
on field SFGs, higher-density environments have received more recent
attention \citep[e.g.,][]{greene2012,jaffe2014,lin2014,erfanianfar2016}. Along those lines, we plot SFR versus stellar mass in the upper panels of Figure\ \ref{fig:mass_SFR},
for the star-forming cluster galaxies (green stars) selected in Section\ \ref{sub:SF_ and_quiescent_galaxies}. Our cluster SFGs are separated into two broad redshift
bins, $0.15<z<0.584$ and $0.584<z<1.5$, each spanning $\sim$3.75 Gyr in lookback time. The galaxies are fitted (dark green lines) with Equation\ (\ref{eq:SFR_M_powerLaw}) and the results are reported in each panel. Despite the considerable scatter in our SFR-$M_{\star}$ fits, the standard deviation of 0.38 at $0.15<z<0.584$ is similar to the observed 1$\sigma$ scatter in the literature, which ranges from 0.15 to 0.61, as compiled by \citet{speagle2014} for studies where the redshift ranges overlap with ours. Similarly, over $0.584<z<1.5$, our 1$\sigma$ scatter of 0.36 is consistent with the corresponding range (0.09 to 0.47) of measured standard deviations reported by \citet{speagle2014}. Shown in the upper panels are 1$\sigma$ uncertainties per parameter derived using the formalism from \citet{press1992}, assuming the uncertainty of each ordinate value is the mean SFR uncertainty. The total uncertainties, shown in parenthesis, account for possible field contamination in the non-spectroscopically selected cluster SFGs. We devise a Monte Carlo realization to calculate the uncertainty due to potential interlopers, whereby 74\% of the $z_{\mathrm{phot}}$/color-selected subset (to simulate a 26\% interloper rate; see Section\ \ref{sub:cluster_membership}) are randomly selected. These galaxies are combined with the $z_{\mathrm{spec}}$ members and a new fit is re-derived. This is performed 5000 times in each redshift bin, and the standard error is calculated for the 5000 sets of parameters. The standard error is added in quadrature with the 1$\sigma$ uncertainty to derive a total parameter uncertainty.

In the lower panels of Figure\ \ref{fig:mass_SFR}, we show our best-fit $\alpha$ and $\beta$, with corresponding uncertainties, along with a sample of best-fit SFR-$M_{\star}$ parameters and uncertainties (if available) from the literature, as calibrated and provided by \citet{speagle2014}. Given the consistency of our fits with those from the literature, this suggests that (even when accounting for any potential contamination from interlopers) the SFR-$M_{\star}$ relation of \textit{star-forming} cluster galaxies is robust and effectively the same as that of field SFGs, echoing similar conclusions by, e.g., \citet{greene2012}, \citet{koyama2013}, and \citet{lin2014}.

\subsection{The Impact of Stellar Mass on Star-forming and Quiescent Cluster
Galaxies}

\label{sub:mass_fQ_sSFR}

\begin{figure}[!t]
\includegraphics{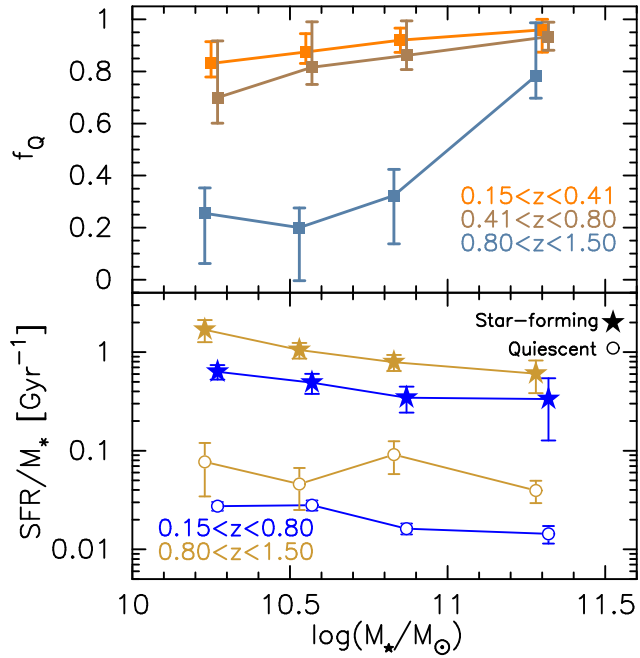}

\caption{Upper panel: Fraction of quiescent cluster galaxies as a function
of stellar mass. Lower panel: Specific SFR versus stellar mass for cluster SFGs (stars)
and quiescent galaxies (open circles). Points have been slightly offset for clarity.\label{fig:mass_fQ_sSFR_z}}
\end{figure}

The upper panel of Figure\ \ref{fig:mass_fQ_sSFR_z} shows the fraction
of quiescent cluster galaxies, $f_{\mathrm{Q}}$, as a function of
stellar mass. Given our stellar mass binning, three cluster galaxies
with $\log\left(M_{\star}/M_{\odot}\right)>11.6$ are excluded. Their
removal plays no role in our final results. In each stellar mass bin,
galaxies are separated into three redshift slices, each spanning $\sim$2.5
Gyr of lookback time (see legend for color coding).

The fraction of quiescent galaxies and stellar mass are correlated,
regardless of redshift. However, the increase in $f_{\mathrm{Q}}$ with stellar mass is relatively mild at $z\sim0.3$ and $z\sim0.6$, and given the uncertainties, formally consistent with no trend. Over $0.2<z<0.5$ and $0.5<z<0.8$, which are similar to our two lower redshift bins,
\citet{lin2014} found $f_{\mathrm{Q}}$ values that are uniformly
lower at $\log\left(M_{\star}/M_{\odot}\right)\lesssim11.3$, even
when accounting for the difference between their adopted \citet{salpeter1955}
IMF and our \citet{chabrier2003} IMF. While \citet{lin2014} did
not provide the halo mass range of their sample, they differentiated
between groups and clusters based on richness. Their cluster selection
corresponds to halo masses $\gtrsim$$10^{14}\,M_{\odot}$, which
is nearly an order of magnitude lower than the minimum $M_{200}$
of our cluster sample over the same redshift range ($\sim$$8\times10^{14}\,M_{\odot}$).
As quiescent galaxies tend to populate more massive halos at a given
stellar mass \citep{lin2016}, the large difference in the underlying
properties of the two cluster samples may explain the substantial
differences in $f_{\mathrm{Q}}$.

The high fraction of quiescent galaxies at $z\lesssim0.8$ implies
that most of the passive cluster population was being built up at
earlier times. Since the most massive cluster galaxies ($\log\left(M_{\star}/M_{\odot}\right)\sim11.3$)
are almost uniformly passive as far back as $z\sim1.2$, this build
up can be attributed to the quenching of lower-mass galaxies at (or
above) this redshift.

Quiescent fraction and stellar mass are more strongly correlated at
$z\sim1.2$ than at $z\sim0.3$ or $z\sim0.6$, with $\Delta f_{\mathrm{Q}}=0.53$
over the same stellar mass range as in the two lower redshift bins.
Qualitatively, this matches the results of \citet{balogh2016} for
galaxies in $\log\left(M_{200}/M_{\odot}\right)\sim14.5$ clusters
at $z\sim1$, although they found that $f_{\mathrm{Q}}$ rises sharply
at $\log\left(M_{\star}/M_{\odot}\right)\sim10.5$, followed by a
gradual increase up to $\log\left(M_{\star}/M_{\odot}\right)\sim11.3$.
The correlation between $f_{\mathrm{Q}}$ and $M_{\star}$ above $z=0.8$
suggests that quenching is ongoing at this epoch and the build up
of stellar mass plays a role in turning off the SF activity of SFGs.
\citet{lee2015} found a strong correlation between quiescent fraction
and stellar mass for candidate cluster galaxies over $1<z<1.5$, which
they also attribute to mass quenching.

Some portion of the increase in quiescent fraction over time at fixed
mass, however, is likely due to the accretion of previously-quenched
galaxies from lower-density environments into clusters \citep[pre-processing; e.g.,][]{haines2015}.
Regardless of the method by which cluster galaxies are quenched, or
even whether they fall into the cluster environment pre-quenched,
it is the increasing quiescent fraction that drives the strong evolution
in the SF activity of the overall cluster population with time (Section\ \ref{sec:z_sSFR}).

In the lower panel of Figure\ \ref{fig:mass_fQ_sSFR_z} we consider
the SF efficiency of our cluster galaxies as a function of stellar
mass. Due to the paucity of low-redshift SFGs, particularly at higher
stellar mass, we combine the two lower redshift bins (see legend). While $f_{\mathrm{Q}}$ shows a strong dependence on $M_{\star}$
at higher redshift, the sSFRs of both SFGs and quiescent galaxies
are more mildly correlated with stellar mass. In fact, at a given
redshift, there is at most a factor of $\sim$$2-3$ difference between
the sSFR of the lowest- and highest-mass cluster galaxies within each
subset. We find a similar trend in the tabulated results of \citet{lin2014},
over the same stellar mass range that we study, and the cluster galaxy
sSFRs measured by \citet{muzzin2012} over $0.85<z<1.2$ align extremely
well with ours at $0.8<z<1.5$. While the build up of stellar mass
appears linked to transitioning galaxies between the star-forming
and quiescent populations, stellar mass has less of an impact on the
SF efficiency of galaxies within each distinct subset.

\subsection{The Evolution of Cluster Galaxies: The Mixing of Two Distinct Populations}

\label{sec:z_sSFR}

\begin{figure}[!t]
\includegraphics{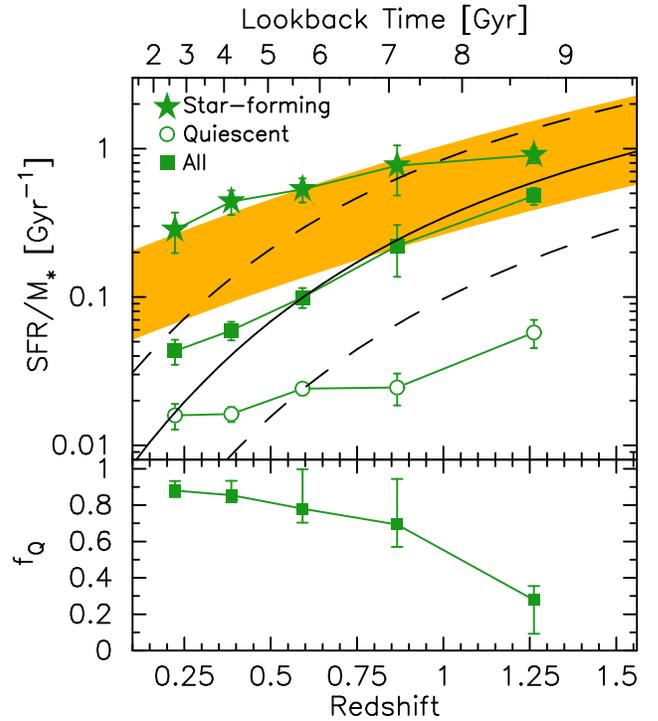}

\caption{Upper panel: Specific SFR versus redshift for different cluster galaxies
(symbols). At all redshifts, the overall cluster population (squares)
is composed of two distinct subsets: quiescent galaxies (open circles)
and SFGs (stars). The sSFR of cluster SFGs is typically consistent
with the SF main sequence from \citet[gold shaded region]{elbaz2011}.
The overall cluster sSFR is in excellent agreement with the \citet{alberts2014}
fit of sSFR versus redshift for cluster galaxies (solid black curve;
dashed curves are 1$\sigma$ uncertainty). Lower panel: Fraction of
quiescent cluster galaxies versus redshift. The quiescent population
builds up quickly at earlier times, with fractions gradually increasing
to more recent times.\label{fig:z_sSFR_fQ}}
\end{figure}

\begin{figure*}[!t]
\begin{centering}
\includegraphics{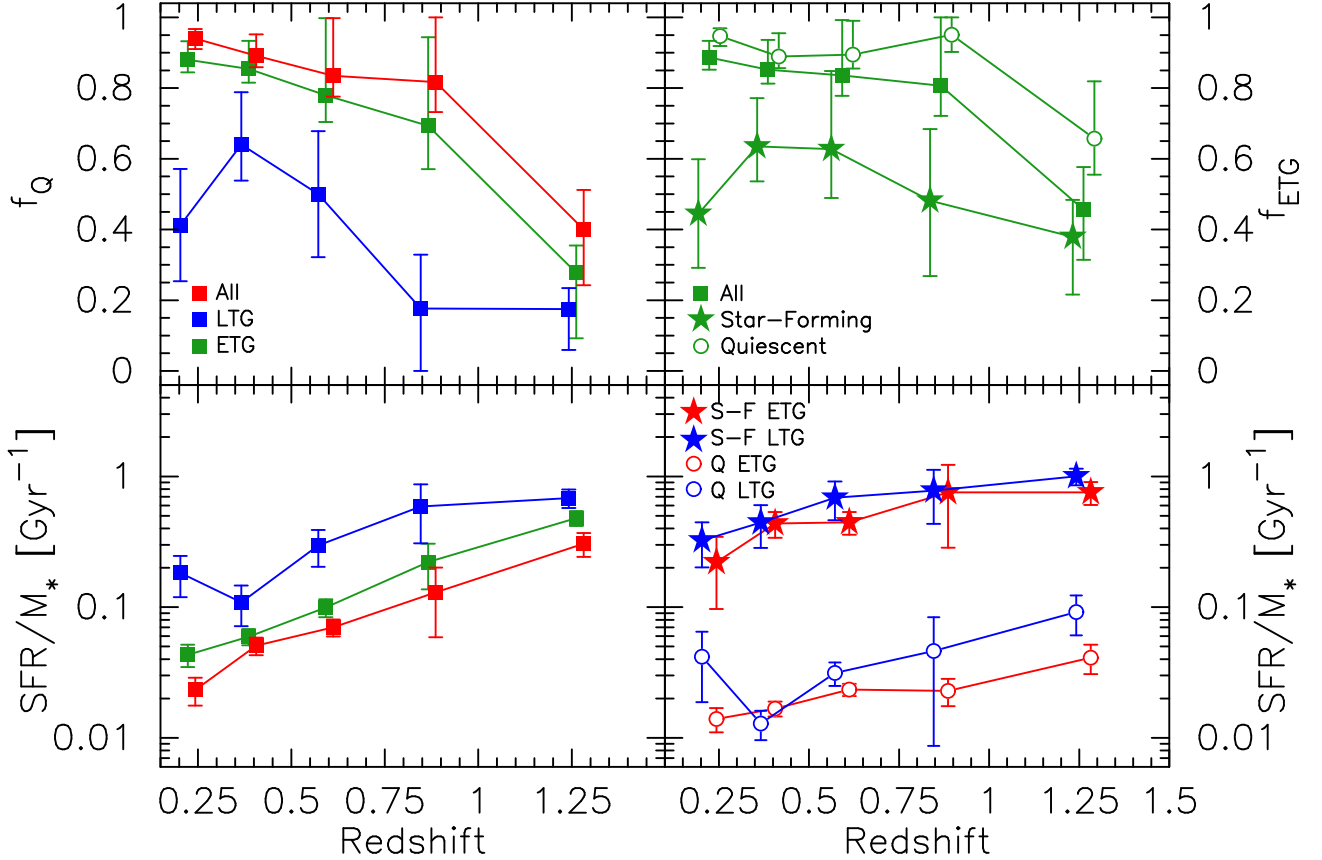}
\par\end{centering}

\caption{Upper left panel: Fraction of quiescent cluster galaxies versus redshift,
separated by morphology. LTGs (blue squares) are predominantly star-forming
at $z\gtrsim0.9$; they become more evenly mixed between SFGs and
quiescent galaxies below this redshift. The quiescent ETG population
(red squares) builds up quickly at earlier times, with fractions gradually
increasing to more recent times. The overall quiescent fraction (green
squares) is well-traced by the ETGs. Upper right panel: Fraction of
ETGs for different subsets of cluster galaxies. While quiescent galaxies
(open circles) and the overall cluster population (squares) are predominantly
early-type at all redshifts, SFGs (stars) have a more mixed morphological
makeup. Lower left panel: Specific SFR versus redshift for all cluster
galaxies (green squares), all ETGs (red squares), and all LTGs (blue
squares). Lower right panel: Specific SFR as a function of redshift
for cluster ETGs (red symbols) and LTGs (blue symbols), separated
into star-forming (``S-F''; stars) and quiescent (``Q''; open
circles) subsets. Independent of morphology, SFGs (and likewise quiescent
cluster members) have very similar SF efficiency at all redshifts.
For clarity, some points in each panel are slightly offset to the
left or right of the bin center.\label{fig:z_fQ_sSFR_morph}}
\end{figure*}

Given that stellar mass has a relatively mild impact on the SF efficiency
of cluster galaxies when independently considering star-forming and
quiescent populations, we now investigate the redshift evolution of
sSFR for these two subsets, along with the overall cluster sample.
The upper panel of Figure\ \ref{fig:z_sSFR_fQ} shows sSFR versus
redshift for all cluster galaxies (green squares), and the star-forming
(green stars) and quiescent (green open circles) subsets, with galaxies
separated into five redshift bins, each spanning $\sim$1.5 Gyr of
lookback time.

For comparison, two sSFR fits from the literature are plotted. \citet{elbaz2011}
measured the sSFR redshift evolution of galaxies observed in the northern
and southern fields of the Great Observatories Origins Deep Survey.
Their best-fit sSFR evolution, which assumes $\alpha=1$ (Equation\ \ref{eq:SFR_M_powerLaw}),
describes a main sequence of
\begin{equation}
13\left(13.8\,\mathrm{Gyr}-t_{L}\right)^{-2.2}\leq sSFR\leq52\left(13.8\,\mathrm{Gyr}-t_{L}\right)^{-2.2},
\end{equation}
which is plotted as the gold shaded region. For a given $t_{\mathrm{L}}$,
\citet{elbaz2011} classified galaxies above this range as starbursts;
galaxies that lie below $13\left(13.8\,\mathrm{Gyr}-t_{L}\right)^{-2.2}$
were considered to have ``significantly lower'' SF.

\citet{alberts2014} modeled the evolution of SF activity in cluster
galaxies over $0.3<z<1.5$, shown here as the solid black curve (dashed
curves are the 1$\sigma$ uncertainties in their fit). Their fit only
includes galaxies with cluster-centric distances of $R_{\mathrm{proj}}<0.5\,\mathrm{Mpc}$;
similarly excluding all galaxies beyond this radius from our sample
would not alter our qualitative results.

At all redshifts, the overall cluster population lies in good agreement
with the \citet{alberts2014} relation, with sSFR decreasing by a
factor of 11 from $z\sim1.3$ ($sSFR=0.48\pm0.06\;\mathrm{Gyr}^{-1}$)
to $z\sim0.2$ ($sSFR=0.043\pm0.009\;\mathrm{Gyr}^{-1}$). The SF
efficiency of the SFG subset also declines over the same period, but
only by a factor of 3. We find that this decline is somewhat shallower
than the best-fit \citet{elbaz2011} sSFR evolution, although our
values are consistent within the uncertainties with their main
sequence. That cluster SFGs have similar SF efficiency to the field
population is not surprising given our results from Section\ \ref{sub:SFR-mass_relation},
where we found that the SFR-$M_{\star}$ relation of cluster and field
SFGs are consistent with each other, despite large scatters about
the best fits. Quiescent cluster galaxies, which have SF efficiency
$\sim$$16-31$ times lower than SFGs, experience a similar decrease
in sSFR to their star-forming counterparts from $z\sim1.3\rightarrow0.2$,
although with a slightly steeper factor of 4. As with SFGs, this evolution
is more moderate than that of the overall cluster population. Thus,
the strong evolution in cluster galaxy sSFR cannot simply be due to
the decline in the sSFR of its constituent subsets, which experience
relatively shallow decreases. Instead, as shown below, the build up
of the quiescent population has a strong impact in the sSFR evolution
of all cluster galaxies.

The lower panel of Figure\ \ref{fig:z_sSFR_fQ} shows $f_{\mathrm{Q}}$
of all cluster galaxies as a function of redshift, using the same
binning as in the top panel. At $z\sim1.3$, the overall quiescent
fraction is only $28^{+8}_{-19}\%$. Based on their relatively small contribution
to the cluster population, quiescent galaxies have little impact on
the overall (high) sSFR at this redshift. After a sharp rise to $f_{\mathrm{Q}}=0.69^{+0.25}_{-0.12}$ at $z\sim0.9$, the quiescent fraction increases gradually with decreasing redshift to
$z\sim0.2$, where $88^{+5}_{-4}\%$ of cluster galaxies are quiescent.
Hence, from $\sim$$7$ to 3 Gyr ago, quiescent galaxies progressively
dominate the cluster population, thus lowering the SF efficiency of
the overall cluster population, more so than their own moderate evolution
would imply.

\subsection{Star-forming and Quiescent Cluster Galaxies: Disentangling Morphology}

We now explore whether the morphologies of different cluster galaxies
play a role in the SF activity of the overall population. The evolution
of the quiescent fraction for cluster ETGs and LTGs is plotted in
the upper left panel of Figure\ \ref{fig:z_fQ_sSFR_morph}, with
each redshift bin spanning $\sim$1.5 Gyr of lookback time. The $f_{\mathrm{Q}}$
of cluster galaxies varies by morphology at all redshifts. While ETGs
quickly build up the majority of their quiescent population at earlier
times ($z\gtrsim0.9$), LTGs are predominantly star-forming at redshifts
above $z\sim0.6$. Furthermore, over $0.15<z<1.5$, the LTG quiescent
fraction is uniformly lower than that of the ETG subset, by up to
a factor of $\sim$5.

As SFGs comprise a larger component of the LTG population, it would
be no surprise to find that the SF activity of \textit{all} cluster
LTGs is higher than that of \textit{all} cluster ETGs (e.g., Paper
I), given the substantial difference in SF efficiency between the
overall quiescent and SFG populations (see upper panel of Figure\ \ref{fig:z_sSFR_fQ}).
Indeed, we find just such a relation for the sSFR of cluster galaxies,
separated by morphology, in the lower left panel of Figure\ \ref{fig:z_fQ_sSFR_morph}.
Across all redshifts, cluster LTGs are $\sim$$2-8$ times more star-forming
than their early-type counterparts. However, as we noted in Section\ \ref{sec:z_sSFR},
simply combining SFGs and quiescent galaxies washes out the strong
differences between these two \textit{distinct} types of galaxies.
Instead, the morphological segregation of galaxies within the SFG
subset (lower right panel of Figure\ \ref{fig:z_fQ_sSFR_morph})
indicates that ETGs and LTGs have remarkably consistent sSFR values.
Quiescent galaxies also show a strong agreement between ETG and LTG
SF efficiency. Star-forming ETGs have SF activity $\sim$$16-33$
times higher than their quiescent counterparts; the differences between
quiescent and star-forming LTGs are similar (factors of $\sim$$8-34$).
At each redshift, the sSFR difference between star-forming and quiescent
galaxies \textit{of the same morphology} is almost uniformly an order
of magnitude or more, while there is at most a minor difference \textit{between
morphologies} within either the SFG or quiescent subsets.

As noted in Section\ \ref{sec:z_sSFR}, the sSFR of the overall cluster
population decreases by a factor of 11 from $z\sim1.3$ to $z\sim0.2$.
This decline is larger than that of LTGs, whose SF efficiency drops
by a factor of 4 (or 6 if only considering the evolution down to $z\sim0.4$).
The stronger overall evolution is due to the build up of the ETG population
with time, as seen in the upper right panel of Figure\ \ref{fig:z_fQ_sSFR_morph},
where we plot $f_{\mathrm{ETG}}$, the fraction of cluster galaxies
classified as ETGs (squares). While $z\sim1.3$ clusters contain a
similar distribution of LTGs and ETGs, at later times, the latter
become---and remain---the dominant galaxy morphology. While the SF
efficiency of ETGs is no different from LTGs when accounting for ongoing
SF, a higher fraction of ETGs are quiescent relative to LTGs. Thus,
the confluence of these facts, combined with the increasing preponderance
of ETGs, contributes to driving the observed cluster sSFR evolution.

The morphological distribution of the star-forming subset is almost uniformly distinct
from that of the cluster population below $z\sim1.3$ (upper right
panel of Figure\ \ref{fig:z_fQ_sSFR_morph}). While the \textit{overall}
cluster sample quickly becomes ETG dominated below $z\sim1.3$, SFGs
experiences a potentially more gradual rise in $f_{\mathrm{ETG}}$
with time and their ETG fraction never exceeds $\sim$$60\%$. Even
though the majority of cluster ETGs are indeed passive across $0.15<z<1.5$,
they still contribute approximately half of the SFG population. At
higher redshifts, even as the entire cluster population acquires more
late-type properties, star-forming ETGs remain prevalent \citep[e.g.,][]{mei2015}.
These results, combined with the sSFR uniformity of \textit{all} cluster
SFGs, suggests that the assumption that most (or all) cluster ETGs
are ``red and dead'' oversimplifies their actual diversity.

\subsection{The Uncertain Path to Quiescence}

Addressing the nature and effectiveness of the quenching mechanisms
at play in our CLASH and ISCS cluster galaxies is clearly challenging,
however, given our results, we can draw some broad inferences. For
instance, if a sizable fraction of cluster galaxies are being slowly
quenched (strangulation), the SF activity of the SFG subset should
(na\"{i}vely) be lower than that of field galaxies \citep[e.g.,][]{paccagnella2016}.
This is supported by \citet{lin2014}, who found that cluster SFGs
over $0.2<z<0.8$ have 17\% lower SF activity than field SFGs at fixed
stellar mass, attributing the difference to strangulation. However,
at all redshifts we study the SF activity of \textit{actively} star-forming
cluster galaxies is roughly equivalent to that of field SFGs (upper
panel of Figure\  \ref{fig:z_sSFR_fQ}). While these binned sSFRs
are consistent with, or even slightly higher than, the field main
sequence of \citet{elbaz2011}, there are some individual SFGs ($13\pm2\%$)
that lie below this range, which may suggest that strangulation acts,
at least on some level, out to at least $z=1.5$ \citep[see also][]{alberts2014}. 

However, given the general uniformity between star-forming cluster
and field sSFRs and the overall cluster galaxy quiescent fraction
that increases by 0.41 from $z\sim1.3\rightarrow0.9$ (a time span
of $\sim$1.5 Gyr), we suggest that if cluster galaxies are being
quenched within the cluster environment, it must be happening relatively
quickly for most of them, at least at higher redshifts. This agrees
with \citet{muzzin2012}, who suggested that the lack of an environmental
correlation between sSFR and $D_{n}\left(4000\right)$ in SFGs implies
a rapid transition from star-forming to quiescence for $z\sim1$ cluster
galaxies. While the exact quenching method remains unsettled, there
is evidence that points towards major-merger induced AGN feedback
acting to quench $z\gtrsim1$ cluster galaxies. Specifically, Paper
I found that the fraction of early-type cluster galaxies increases
over $z\sim1.4\rightarrow1.25$, and previous studies have found evidence
for rapid mass assembly in $z\gtrsim1$ clusters \citep{mancone2010,fassbender2014},
young galaxies continuously migrating onto the cluster red sequence
\citep{snyder2012}, and substantial (in some cases field-level) SF
activity \citep{brodwin2013,zeimann2013,alberts2014,webb2015}. These
results are all indicative of ongoing major-merger activity. Furthermore,
the incidence of AGNs in $z\gtrsim1$ ISCS clusters is increased relative
to lower redshifts \citep{galametz2009,martini2013,alberts2016}.

\section{Summary}

\label{sec:conclusion}

We have combined two galaxy samples of varying redshift to conduct
a study of the evolution of cluster galaxy SF activity over $0.15<z<1.5$.
Our final sample contains 11 high-redshift ($1<z<1.5$) ISCS clusters
previously studied in Paper I \citep{wagner2015}, and 25 low-redshift
($0.15<z<1.0$) CLASH clusters. Physical galaxy properties (i.e.,
SFRs and stellar masses) were measured through broadband SED fitting
with CIGALE. Star-forming and quiescent cluster members were separated
based on their CIGALE-derived $D_{n}\left(4000\right)$ values, and
galaxies were classified into ETGs and LTGs based on their morphologies
through visual inspection of high-resolution \hst\ images.

Cluster LTGs were found to have uniformly higher SF activity than
ETGs, though this is caused by an increase in the fraction of quiescent
ETGs with time. At each considered redshift, star-forming ETGs have
an sSFR consistent with that of late-type SFGs.

From $z\sim1.3$ to $z\sim0.2$, the sSFR of cluster SFGs declines
by a relatively modest factor of 3. Their quiescent counterparts experience
a similar (factor of 4) decrease in SF efficiency, while maintaining
sSFRs at least an order of magnitude lower than SFGs. The evolution
of the overall cluster sSFR, which agrees well with \citet{alberts2014},
is largely driven by the relative fraction of its constituent populations.
With their sSFRs matching the field main sequence from \citet{elbaz2011},
cluster SFGs provide the dominant contribution to the overall population
at higher redshifts ($z\sim1.3$), where they comprise more than $70\%$
of the cluster population. However, as the fraction of quiescent cluster
galaxies rises with decreasing redshift, up to $f_{\mathrm{Q}}=0.88^{+0.05}_{-0.04}$
at $z\sim0.2$, their relatively low sSFRs have a much greater impact
on the overall SF activity. This culminates in a factor of 11 decrease
in sSFR for all massive cluster galaxies from $z\sim1.3$ to $z\sim0.2$.

By comparing the sSFRs of our cluster SFGs with the main sequence
of \citet{elbaz2011}, we found a subset with low field-relative SF
activity, which makes up $13$\% of the cluster SFG population. This
is indicative of strangulation acting on some level in our clusters.
However, the approximately field-level SF activity of cluster SFGs
and the quiescent fraction increasing by 0.41 from $z\sim1.3\rightarrow0.9$
would suggest that at $z\gtrsim0.9$, the mechanism(s) quenching cluster
galaxies is likely a rapid process (e.g., merger-driven AGN quenching). 

While our results provide additional evidence for the rapid quenching
of higher-redshift cluster galaxies, much uncertainty remains in constraining
the dominant quenching mechanisms. Further study is required to better,
and more quantitatively, measure the contributions of the various
quenching mechanisms.

\acknowledgements

We appreciate the numerous helpful suggestions provided by the referee. We are  grateful to D. Lemze for providing his spectroscopic membership list for Abell 611 in digital form, and to the CLASH collaboration for making their data publicly available. We thank M. Balogh, D. Marchesini, and M. McDonald for constructive discussions, P. Eisenhardt for a helpful review of the paper, and A. Karunakaran for compiling a thorough list of cluster velocity dispersions. CW acknowledges support from the estate of Urlla Eillene Carmichael through a generous Duncan and Urlla Carmichael Fellowship. SC acknowledges support of the Natural Sciences and Engineering Research Council of Canada through a generous Discovery grant. GS acknowledges support from HST grant HST-AR-13887.004-A. The work of DS was carried out at Jet Propulsion Laboratory, California Institute of Technology, under a contract with NASA. Support for Program \#13887 was provided by NASA through a grant from the Space Telescope Science Institute, which is operated by the Association of Universities for Research in Astronomy, Inc, under NASA contract NAS5-26555. This research made use of NASA's Astrophysics Data System, Ned Wright's online cosmology calculator \citep{wright2006}, and \textsc{topcat}, an interactive graphical viewer and editor for tabular data \citep{taylor2005}.

\appendix

\section{Comparison of CIGALE-derived Star Formation Rates and Stellar Masses}

\label{sec:SFR_mass_comparison}

\begin{figure}[!t]
\begin{centering}
\includegraphics[width=\columnwidth]{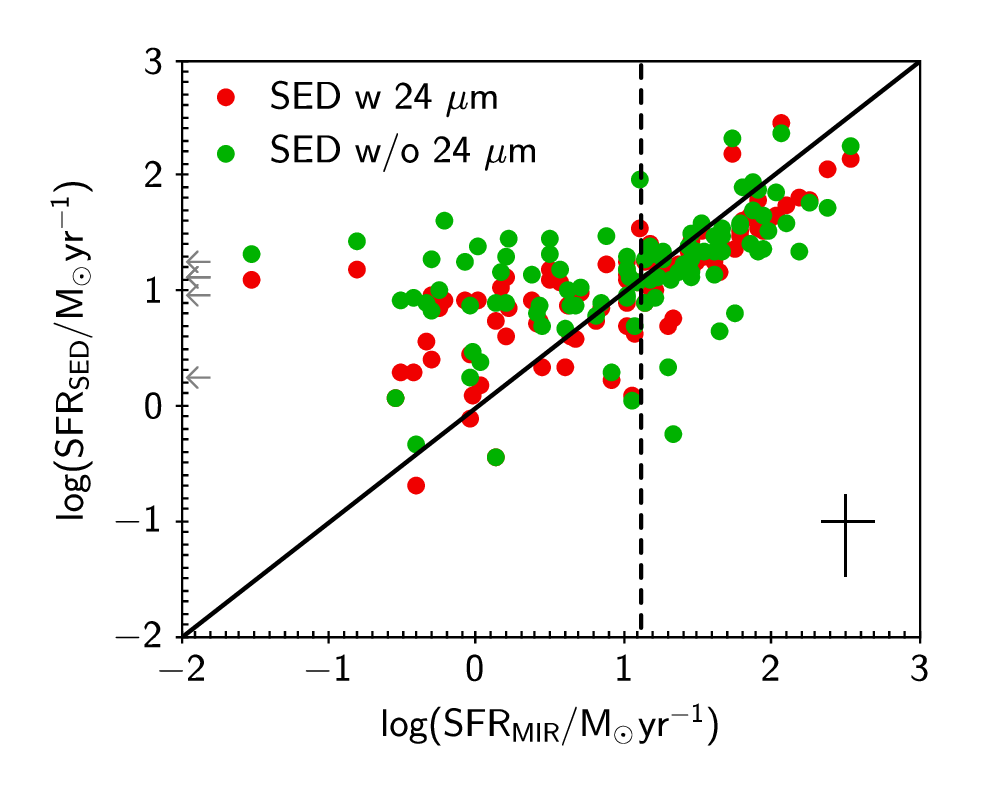}
\par\end{centering}

\begin{centering}
\includegraphics[width=\columnwidth]{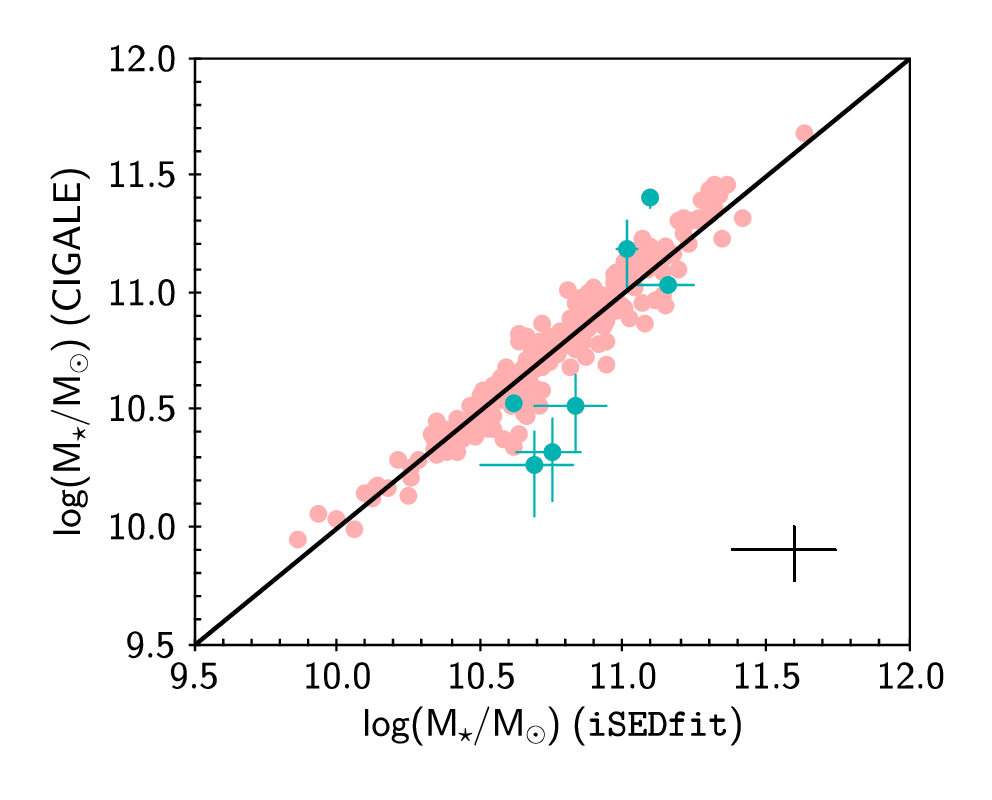}
\par\end{centering}

\caption{Upper panel: Comparison of SFRs derived through SED fitting ($SFR_{\mathrm{SED}}$)
and 24 $\mu$m SFRs calculated in Paper I ($SFR_{\mathrm{MIR}}$).
Galaxies plotted in red include mid-infrared fluxes in the SED fits,
while those plotted in green do not. The gray arrows are the $SFR_{\mathrm{SED}}$
of galaxies undetected at 24 $\mu$m. The vertical dashed line shows
the 1$\sigma$ $SFR_{\mathrm{MIR}}$ limit of $\sim$$13\,M_{\odot}\,\mathrm{yr}^{-1}$.
Lower panel: Comparison of ISCS stellar masses derived with CIGALE
and with \texttt{iSEDfit}. The galaxies highlighted in turquoise are
the seven where the two measurements do not agree within the uncertainties.
In both panels the solid line is a 1:1 relation, while the black error
bars show the typical uncertainties in the derived properties.\label{fig:ISCS_SFR_mass_comp}}
\end{figure}

The upper panel of Figure\ \ref{fig:ISCS_SFR_mass_comp} shows the
SFRs of isolated ISCS galaxies as measured by CIGALE ($SFR_{\mathrm{SED}}$)
versus those calculated in Paper I ($SFR_{\mathrm{MIR}}$). We run
CIGALE twice with the same parameters, first with 24 $\mu$m flux
points, and then without. We compare $SFR_{\mathrm{SED}}$ from these
runs against $SFR_{\mathrm{MIR}}$ from Paper I with red and green
points, respectively. The dashed line shows the 1$\sigma$ SFR limit
of $\sim$$13\,M_{\odot}\,\mathrm{yr}^{-1}$.

We find the median absolute deviation (MAD) of
\begin{equation}
\Delta\log\left(SFR\right)=\log\left(\frac{SFR_{\mathrm{MIR}}}{M_{\odot}\,\mathrm{yr}^{-1}}\right)-\log\left(\frac{SFR_{\mathrm{SED}}}{M_{\odot}\,\mathrm{yr}^{-1}}\right)\label{eq:SFR_MAD}
\end{equation}
 for all galaxies plotted here, both including and excluding the 24
$\mu$m fluxes, and list them in Table\ \ref{tab:SFRandMassComps}.
We similarly list the MAD for the SFR differences when only considering
the galaxies above the 1$\sigma$ SFR limit.

\begin{deluxetable}{ccc}
\tablecaption{MAD of $\Delta\log\left(SFR\right)$ Between Paper I and CIGALE\label{tab:SFRandMassComps}}
\tablewidth{0pt}
\tablehead{
\colhead{24 $\mu$m Flux in SED} &
\colhead{$SFR_{\mathrm{MIR}}$ Cut} &
\colhead{MAD}
\\
\colhead{} &
\colhead{($M_{\odot}\,\mathrm{yr}^{-1}$)} &
\colhead{}
}
\startdata
Yes & $>$0 & 0.26 \\
Yes & $>$13 & 0.11 \\
No & $>$0 & 0.33 \\
No & $>$13 & 0.16 \\
\enddata
\end{deluxetable}

When considering galaxies with low levels of 24 $\mu$m SFR, where
the uncertainty in the mid-infrared is high, both CIGALE runs fail
to accurately reproduce $SFR_{\mathrm{MIR}}$, as can be seen by the
relatively high MADs. However, above the 1$\sigma$ level of $SFR_{\mathrm{MIR}}$,
both sets of $SFR_{\mathrm{SED}}$ compare favorably, with similar
scatter about the 1:1 line, and MADs that decrease by $\sim$0.15.
Given that the $SFR_{\mathrm{SED}}$ values calculated with and without
mid-infrared fluxes are similar, we opt to use the latter in our analysis.
This provides the additional benefit of allowing us to include non-isolated
ISCS galaxies in our analysis, as our SED fits are not contingent
on observations with poor resolution inherent to the mid-infrared.
By including non-isolated ISCS galaxies, the $z>1$ portion of our
sample has 255 galaxies after all selection cuts are applied, $\sim$2.5
times as many as it would have had if only allowing isolated galaxies.

The lower panel of Figure\ \ref{fig:ISCS_SFR_mass_comp} shows stellar
masses derived by CIGALE versus those derived by \texttt{iSEDfit}.
The salmon colored points are the galaxies where the two values agree
within their uncertainties, while the turquoise points show the seven
galaxies where the stellar masses do not agree. Overall, CIGALE and
\textsc{iSEDfit} agree well with each other. We find a MAD of $\Delta\log\left(M_{\star}\right)=0.05$
between the stellar masses derived with the two codes. We calculate
a Pearson correlation coefficient of $r=0.96$ between the CIGALE
and \textsc{iSEDfit} stellar masses.

\bibliographystyle{aasjournal}
\bibliography{Wagner16}

\end{document}